\documentclass[twocolumn, secnumaRabic, amssymb, nobibnotes, aps, showpacs,superscriptaddress,longbibliography]{revtex4-1}
\usepackage{graphicx,subfigure,amsmath}%
\usepackage{color}
\usepackage[colorlinks=true, linkcolor=blue,urlcolor=blue,anchorcolor=blue,citecolor=blue,bookmarksnumbered]{hyperref}

\usepackage{amsmath,amstext}
\usepackage{graphicx}
\usepackage{booktabs}
\usepackage{multirow}
\definecolor{green}{rgb}{0.8,0.98,0.83}
\usepackage{float}
\usepackage{ulem}
\begin{document}
	\title{Quantum computation via Floquet-tailored Rydberg interactions}
	\author{Jun Wu}
	\affiliation{School of Physics and Optoelectronics Engineering, Anhui University, Hefei 230601,  People's Republic of China}
    \author{Jin-Lei Wu}\email[]{jlwu517@zzu.edu.cn}
	\affiliation{School of Physics, Zhengzhou University, Zhengzhou, 450001, People's Republic of China}
	\author{Fu-Qiang Guo}
	\affiliation{Center for Quantum Sciences and School of Physics, Northeast Normal University, Changchun 130024, People's Republic of China}
	\author{Bing-Bing Liu}
	\affiliation{School of Physics, Zhengzhou University, Zhengzhou, 450001, People's Republic of China}
	\author{Shi-Lei Su}\email[]{slsu@zzu.edu.cn}
	\affiliation{School of Physics, Zhengzhou University, Zhengzhou, 450001, People's Republic of China}
	\affiliation{Institute of Quantum Materials and Physics, Henan Academy of Science, Henan 450046, People's Republic of China}
	\author{Xue-Ke Song}\email[]{songxk@ahu.edu.cn}
	\affiliation{School of Physics and Optoelectronics Engineering, Anhui University, Hefei 230601,  People's Republic of China}
	\author{Liu Ye}
	\affiliation{School of Physics and Optoelectronics Engineering, Anhui University, Hefei 230601,  People's Republic of China}
	\author{Dong Wang}\email[]{dwang@ahu.edu.cn}
	\affiliation{School of Physics and Optoelectronics Engineering, Anhui University, Hefei 230601,  People's Republic of China}
	
	\date{\today}
	\begin{abstract}
		Rydberg atoms stand out as a highly promising platform for realizing quantum computation with significant advantages in constructing high-fidelity quantum gates. Floquet frequency modulation~(FFM), in Rydberg-atom systems, provides a unique platform for achieving precise quantum control and uncovering exotic physical phenomena, paving the way for innovative methodologies in quantum dynamics research. This work introduces a method to realize controlled arbitrary phase gates in Rydberg atoms by manipulating system dynamics using FFM. Notably, this method eliminates the need for laser addressing of individual atoms, significantly enhancing convenience for future practical applications. Furthermore, this approach can be integrated with soft quantum control strategies to enhance the fidelity and robustness of the resultant controlled-phase gates. Finally, as an example, this methodology is applied in Grover-Long algorithm to search target items with zero failure rate, demonstrating its substantial significance for future quantum information processing applications. This work leveraging Rydberg atoms and Floquet frequency modulation may herald a new era of scalable and reliable quantum computing.
	\end{abstract}
	\maketitle
	\section{INTRODUCTION}
	
    Due to unique advantages, neutral atoms have emerged as one of the most promising and rapidly developing platforms in quantum computation and quantum many-body physics~\cite{ PhysRevLett.85.2208,Gallagher_1994,RevModPhys.82.2313,PhysRevLett.110.263201,doi:10.1126/science.aax9743, PhysRevLett.123.170503,Evered2023,PhysRevLett.133.243601}. When excited to Rydberg states, neutral atoms exhibit relatively long lifetimes and strong Rydberg-Rydberg interaction (RRI)~\cite{Anand2024,PhysRevApplied.20.034019,PhysRevA.109.012619,Wu:23,Liu2024}, which can manifest as the form of dipole-dipole or van der Waals forces. This interaction gives rise to the Rydberg blockade mechanism, wherein the excitation of one atom to the Rydberg state prevents the excitation of neighboring atoms~\cite{Gallagher_1994,PhysRevLett.87.037901,Urban2009,PhysRevLett.123.230501,PRXQuantum.4.020335,PhysRevA.109.012615,Sun2024,PhysRevA.109.042621}. This phenomenon has been extensively utilized for the implementation of quantum logic gates~\cite{PhysRevLett.123.170503,PhysRevLett.104.010503,PhysRevApplied.19.044007,PhysRevLett.100.170504,PhysRevResearch.2.043130,PhysRevA.110.032619,PhysRevA.102.062410,PhysRevA.107.062609,PRXQuantum.4.020336,PhysRevA.109.022613} and holds significant promise for various applications in quantum computing, with ongoing experimental advancements~\cite{Cao2024,Jia2024,PhysRevLett.132.150606}. Furthermore, this feature allows quantum information to be encoded in the collective states of atomic ensembles, enabling the realization of mesoscopic quantum information processing and thus forming Rydberg superatoms~\cite{PhysRevLett.121.103601,PhysRevX.5.031015,Shao2024}. Beyond the Rydberg blockade mechanism, Rydberg anti-blockade (RAB)~\cite{Ates2007Antiblockade, PhysRevLett.111.033606,PhysRevLett.111.033607,PhysRevLett.104.013001,PhysRevLett.128.013603,PhysRevA.110.043304,PhysRevA.109.012608} constitutes a critical dynamical process in neutral atomic systems, enabling the simultaneous excitation of multiple atoms to strongly interacting states. In particular, RAB permits a resonant two-photon transition between ground-state pairs and doubly excited Rydberg states in diatomic systems, thus playing a pivotal role in the realization of two-atom phase gates~\cite{PhysRevLett.124.033603,Wu:20,PhysRevA.103.012601} and the preparation of steady-state entanglement~\cite{PhysRevLett.111.033606,PhysRevLett.111.033607,PhysRevA.95.022317}.

    In the realm of coherent quantum dynamics control, periodic driving serves as a prevalent method for state manipulation~\cite{RevModPhys.89.011004,PhysRevLett.120.123204,PhysRevA.103.023335,Nguyen2024,Zhou2024}. By applying Floquet frequency modulation (FFM), which involves periodic modulation of quantum systems at elevated frequencies, dynamic stability can be achieved~\cite{Zhao2023,PhysRevApplied.21.024035,Sun_2024}. Furthermore, through the judicious selection of appropriate modulation amplitude and frequency, the Rabi coupling strength can be effectively regulated. This significantly enhances the dynamics of quantum systems, providing a broader array of choices for manipulating quantum systems. By periodically modulating the atom-field detuning, FFM provides a robust approach to realizing RAB dynamics, irrespective of the strength of the RRI. This technique significantly enhances RAB processes and enables the stabilization of long-lived Rydberg states with strong interactions, even for closely spaced atoms.
	
	The practical execution of quantum computation necessitates the efficient and resilient deployment of quantum logic gates~\cite{PhysRevLett.123.100501,Li2017,Ma2023,Song_2016,PhysRevA.111.012604} to enable the completion of numerous quantum tasks. Nevertheless, most of existing dynamic protocols often fall short of meeting this demand. To address this issue, we employ the FFM method for constructing quantum logic gates. The versatile applications of FFM in coherent quantum system control allow for the exploration of diverse quantum phenomena. Through the manipulation of frequency-modulation parameters of driving field, the system dynamics can be adeptly regulated to realize robust quantum gates. Here we employ FFM to periodically modulate the atom-field detuning, thereby overcoming the constraints imposed by Rydberg anti-blockade on atomic separations. Unlike conventional approaches, which require individual laser addressing of atomic pairs to implement controlled-phase (C-Phase) gates~\cite{PhysRevA.103.012601}, the proposed protocol facilitates the realization of a universal C-Phase gate without the atomic addressing, which significantly enhances the experimental feasibility and operational efficiency. Simultaneously, owing to the intricate dynamics inherent in the Floquet system, varied modulation parameters can be selected to individually achieve the universal C-Phase gate. This offers a broader array of options for the construction of quantum logic gates. Furthermore, the constructed C-Phase gate can be integrated with Gaussian soft quantum control optimization techniques~\cite{PhysRevLett.121.050402,Wu2021,Yin2021GaussianSC,PhysRevA.93.052324,PhysRevApplied.13.044021,PhysRevA.103.032402,Wu2024,PhysRevA.110.043510,PhysRevA.111.022420}, thereby improving the fidelity of the C-Phase gate and mitigating the occurrence of undesirable high-frequency oscillations during the evolution process.
	
	Finally, as an important example of application, we turn to apply our Floquet Rydberg-atom C-Phase gates for implementing Grover-Long search algorithm~\cite{PhysRevLett.79.325,PhysRevLett.80.4329,PhysRevA.64.022307,LONG199927,LONG2002143,He:23,Sankar2024,Li2023,Li2024resourceefficient,Pokharel2024}. By carrying out the phase rotation by an arbitrary angle, Grover-Long algorithm enables the target items search with high success-rate to be realized effectively~\cite{PhysRevA.64.022307}. Our approach facilitates multi-item searching with a unit success probability, eliminating the necessity for individual addressing and simplification of the quantum circuit offers a promising and feasible avenue for the practical implementation of the algorithm. We anticipate that the integration of FFM with advanced quantum control of Rydberg atoms could open new avenues for efficient and robust quantum computation, paving the way for practical implementations of complex quantum algorithms and the exploration of novel quantum phenomena in coherently controlled systems.

	\section{MODEL AND Hamiltonian}\label{sec2}

\begin{figure}
		\centering
		\includegraphics[width=8cm,height=6.86cm]{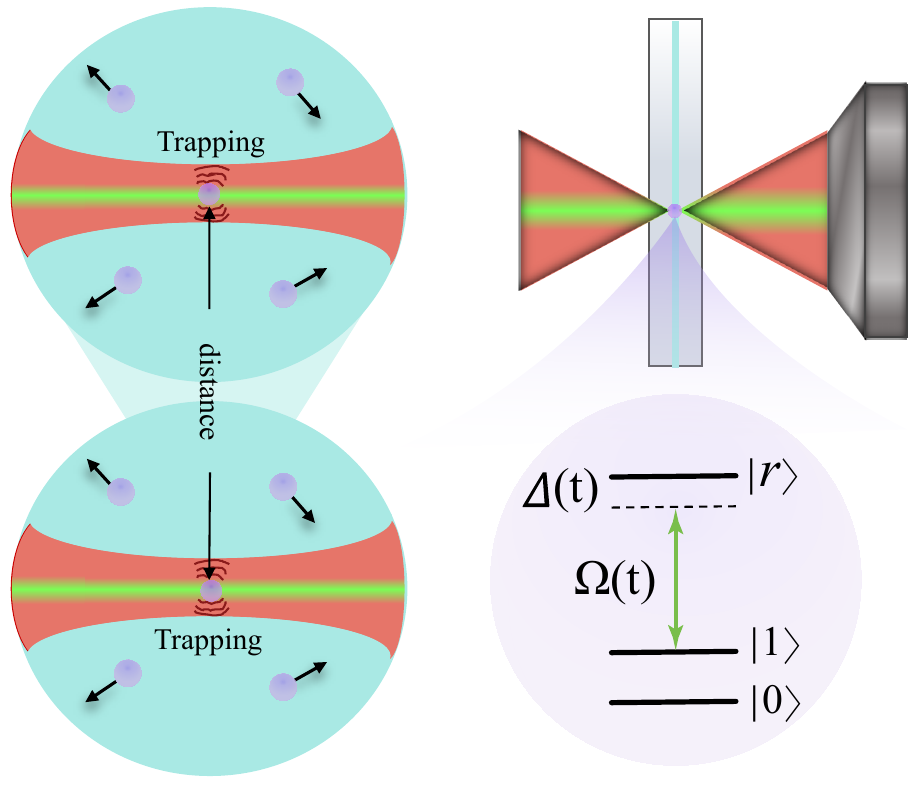}
		\caption{The schematic illustration depicts the energy level configuration of two Rydberg atoms trapped in optical tweezers. Each atom comprises two hyperfine ground states, denoted as $|0\rangle$ and $|1\rangle$, alongside a Rydberg excited state represented by $|r\rangle$. Simultaneously, both atoms undergo excitation from state $|1\rangle$ to state $|r\rangle$ with Rabi frequency $\Omega(t)e^{i\phi(t)}$ and detuning $\Delta(t)$, while exhibiting Rydberg interactions between their respective $|r\rangle$ states.}\label{fig1}
	\end{figure}

	As illustrated in Fig.~\ref{fig1}, we consider a scenario involving two neutral atoms interacting via RRI. Each atom is stimulated from the hyperfine ground state $|1\rangle$ to the Rydberg excited state $|r\rangle$ with Rabi frequency denoted as $\Omega(t)e^{i\phi(t)}$ and detuning $\Delta(t)$. In the interaction picture with rotating-wave approximation~(RWA), the evolution of the system is governed by the Hamiltonian ($\hbar\equiv1$).
	\begin{eqnarray}\label{eq1}
		\hat{H}(t)&=&-\Delta(t)\sum_{i=1}^2|r\rangle_{i}\langle r|+\frac{\Omega(t)}{2}e^{i\phi(t)}\sum_{i=1}^2(|1\rangle_{i}\langle r|+\mathrm{H.c.}) \cr\cr
		&&+V|rr\rangle\langle rr|,
	\end{eqnarray}
    where $i$ indexes the $i$-th atom, $V$ represents the RRI strength of the Van der Waals type, and the laser detuning $\Delta(t)=\Delta_{0}+\delta\sin(\omega_{0}t)$ undergoes sinusoidal FFM with the modulation frequency $\omega_{0}$ and the modulation amplitude $\delta$. The notation $|ab\rangle\equiv|a\rangle_1\otimes|b\rangle_2$ is employed throughout this study for conciseness. In implementing FFM, we can utilize a Rydberg excitation laser controlled by an acoustic-optic modulator (AOM) driven by an arbitrary waveform generator (AWG). Unlike conventional Floquet engineering techniques where FFM does not rely on periodic pulses but it directly alters the effective coupling between energy levels~\cite{PhysRevLett.124.063601,doi:10.1126/science.abd9547}.
    
    Now we introduce the unitary operator, $\hat{U}(t)=\exp\left[-if(t)\sum_{i=1}^2|r\rangle_{i}\langle r|-iVt|rr\rangle\langle rr|\right]$ with $f(t)=-\Delta_0t+\delta/\omega_0\cos(\omega_0t)$, the Hamiltonian~\eqref{eq1} is transformed into $\hat{H}^{\prime}(t)=i\dot{\hat{U}}^{\dagger}(t)\hat{U}(t)+{\hat{U}}^{\dagger}(t)\hat{H}\hat{U}(t)$, calculated as
    \begin{eqnarray}
    	\hat{H}^{\prime}(t)&=&\frac{\Omega(t)}{2}e^{i[\phi(t)+f(t)]}\Big[(\sqrt2|W\rangle\langle 11|+ |0r\rangle\langle 01|\nonumber\\&&
         +|r0\rangle\langle 10|)+\sqrt2e^{iVt}|rr\rangle\langle W|\Big]+\mathrm{H.c.},
    \end{eqnarray}
    in which the single-excitation intermediate state is denoted as $|W\rangle= (|1r\rangle+|r1\rangle)/\sqrt{2}$. We define the modulation index $\alpha=\delta/\omega_0$ and assume $\Delta_{0}=0$ for simplicity. By employing the Jacobi-Anger expansion $\exp[\pm i\alpha\cos\omega_0 t] = \sum_{m=-\infty}^{\infty}J_{m}(\alpha)\exp[\pm im(\omega_0 t+\pi/2)]$, the modified Hamiltonian $\hat{H}^{\prime}(t)$ can be expressed as
    \begin{eqnarray}
    	\hat{H}^{\prime}(t)&=&\frac{\Omega(t)}{2}e^{i\phi(t)}\sum_{m=-\infty}^{\infty}J_{m}(\alpha)e^{im\omega_{0}t+im\frac{\pi}{2}}\Big[(|0r\rangle\langle 01|\cr\cr
    	&&+|r0\rangle\langle 10|+\sqrt2|W\rangle\langle 11|)+\sqrt2e^{iVt}|rr\rangle\langle W| \Big]\cr\cr
    	&&
    	+\mathrm{H.c.},
    \end{eqnarray}
    where $J_{m}(\alpha)$ represents the $m$th order Bessel function of the first kind. To simplify subsequent analyses, the Rabi frequencies between the states are individually rescaled to
    \begin{eqnarray}\label{eq4}
    	\Omega_{a}(t)&=&\Omega(t)e^{i\phi(t)}\sum_{m=-\infty}^{\infty}J_{m}(\alpha)e^{im\omega_{0}t+im\frac{\pi}{2}},
    	\cr\cr\Omega_b(t)&=&\Omega(t)e^{i\phi(t)}\sum_{m=-\infty}^\infty J_m(\alpha)e^{i(m\omega_0+V)t+im\frac{\pi}{2}}.
    \end{eqnarray}
  By selecting a significantly high modulation frequency $\omega_{0}$, the Rabi frequencies $\Omega_{a}(t)$ and $\Omega_{b}(t)$ are predominantly influenced by the resonance terms in Eq.~\eqref{eq4}. Specifically, the Rabi frequency $\Omega_{a}(t)$ is governed by $J_{0}(\alpha)$, whereas for $\Omega_{b}(t)$, we establish $m\omega_{0}=-V$ to satisfy the resonance criterion.
	
		\begin{figure}
		\centering
		\includegraphics[width=\linewidth]{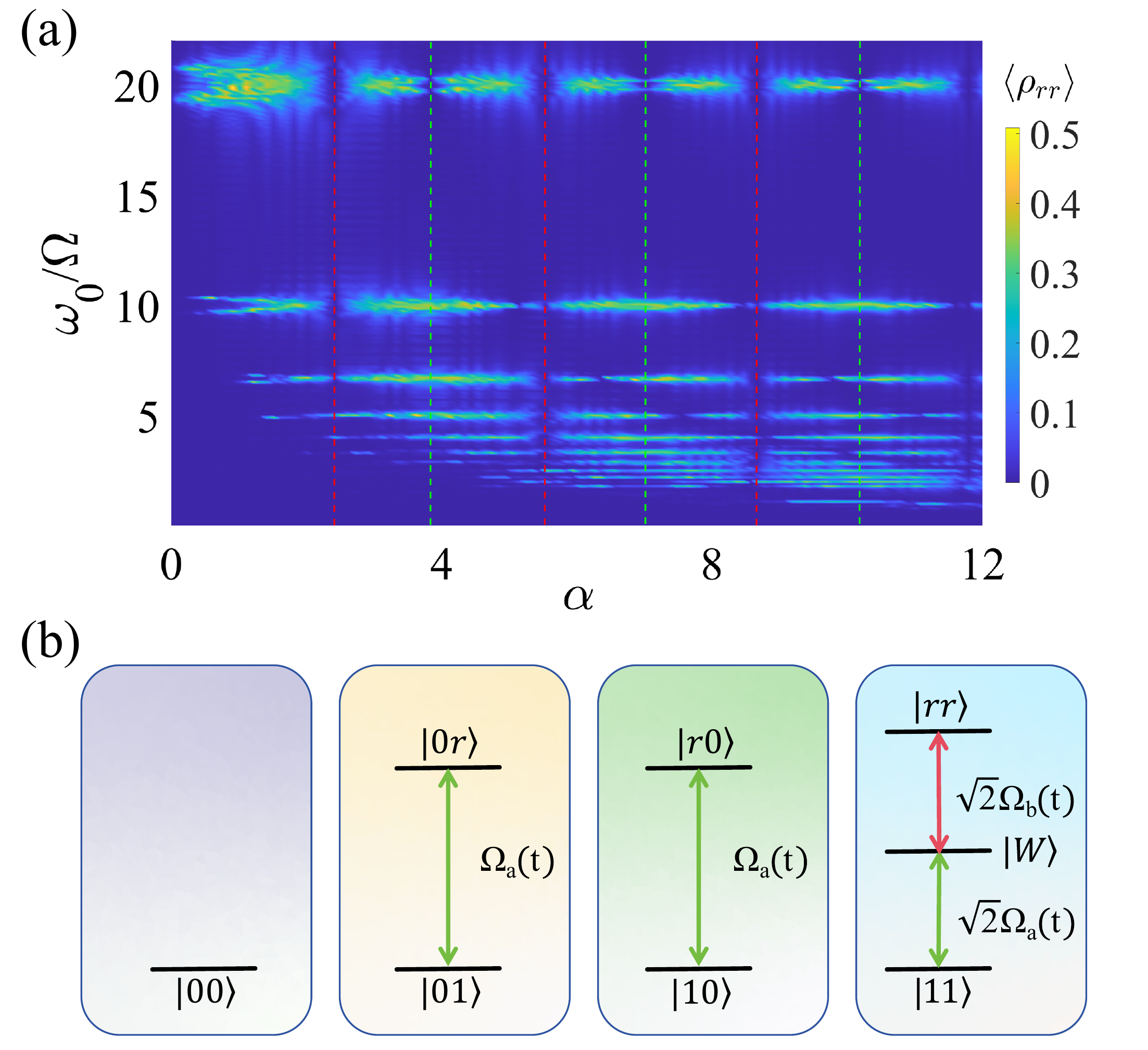}
		\caption{(a) Schematic illustrating the impact of FFM on Rydberg anti-blockade dynamics. The time-average population of $|rr\rangle$ varies as a function of the modulation index $\alpha$ and the normalized modulation frequency $\omega_0/\Omega$. The red and green dashed curves correspond to $J_{0}(\alpha) = 0$ and $J_{1}(\alpha) = 0$, respectively. (b) Dynamics of transitions among the four ground states: the $|00\rangle$ state remains unaltered, transitions of the $|01\rangle$ and $|10\rangle$ states exhibit Rabi oscillations with the same Rabi frequency $\Omega_{a}(t)$, while the $|11\rangle$ state is excited to the $|rr\rangle$ state via the intermediate state $|W\rangle$ with equivalent Rabi frequencies of $\sqrt2\Omega_{a}(t)$ and $\sqrt2\Omega_{b}(t)$, respectively.}\label{fig2}
	\end{figure}
	For a more intuitive analysis, we partition the system dynamics into distinct sectors, specifically involving the computational states $|00\rangle$, $|01\rangle$, $|10\rangle$, and $|11\rangle$. The state $|00\rangle$ remains unchanged as $|0\rangle$ is decoupled to the driving field. In the cases of the initial states $|01\rangle$ and $|10\rangle$, the dynamics can be described in the two-level systems, respectively, governed by
	\begin{eqnarray}
		\hat{H}_{01}(t)&=&\frac{\Omega_a(t)}{2}|01\rangle\langle 0r| + \mathrm{H.c.},
		\cr\cr\hat{H}_{10}(t)&=&\frac{\Omega_a(t)}{2}|10\rangle\langle r0| + \mathrm{H.c.}
	\end{eqnarray}
	For the initial state $|11\rangle$, however, it is reduced into a three-level system with the Hamiltonian
	\begin{equation}\label{eq6}
		\hat{H}_{11}(t)=\frac{\sqrt{2}\Omega_a(t)}{2}|11\rangle\langle W| +\frac{\sqrt{2}\Omega_b(t)}{2}|W\rangle\langle rr| + \mathrm{H.c.}
	\end{equation}
The quite difference between Hamiltonian expressions in the cases of initial states $|01\rangle$~($|10\rangle$) and $|11\rangle$ is the very essence to achieve nontrivial two-atom quantum gates.

        \begin{figure}[b]
	\centering
	\includegraphics[width=0.7\linewidth]{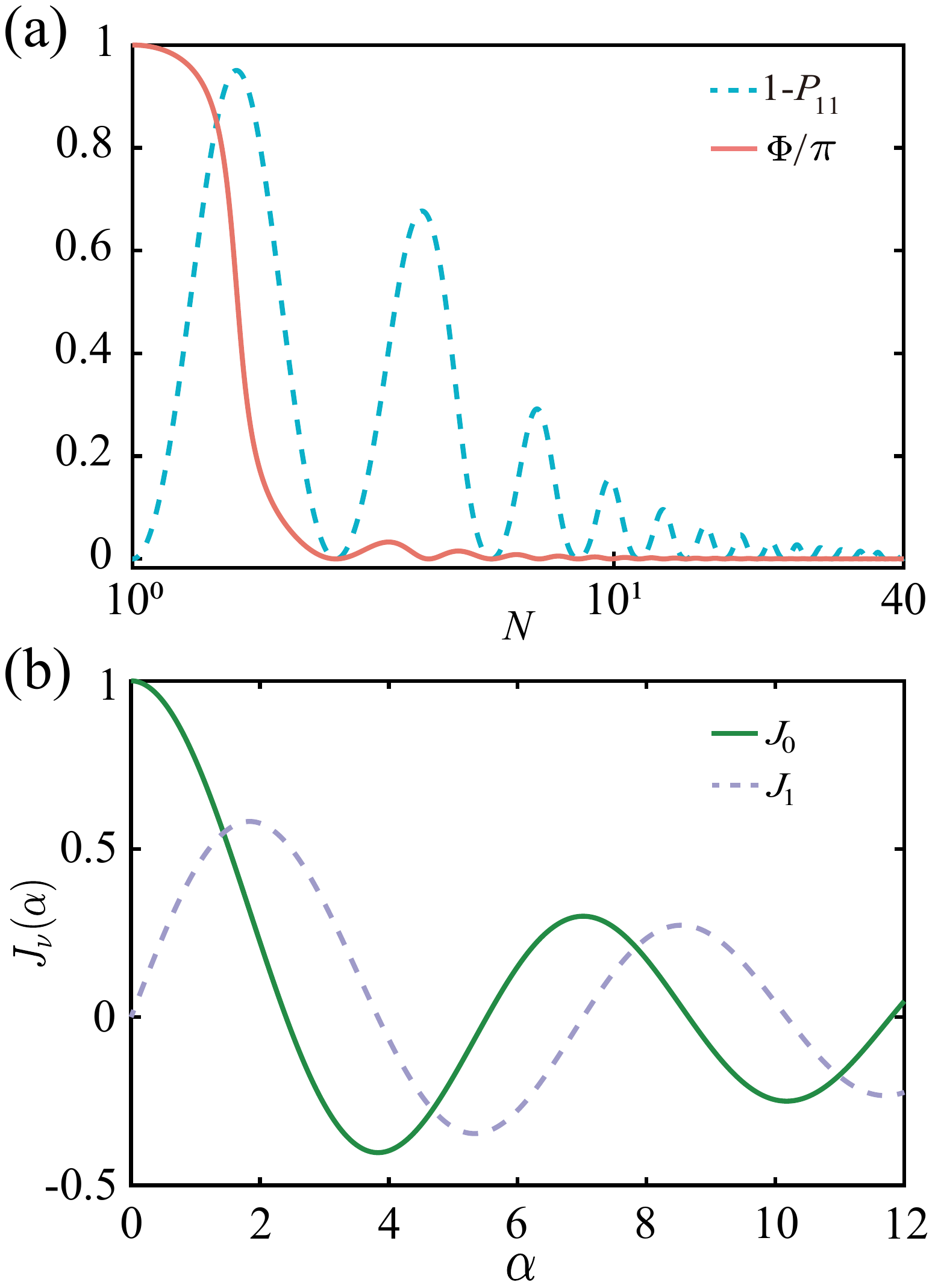}
	\caption{(a) Diagram illustrating the variations in population and phase of $|11\rangle$ as the ratio $N$ of $|\Omega_b(t)|$ to $|\Omega_a(t)|$ is altered. (b) Evolution of the Bessel function of the first kind with respect to the modulation index $\alpha$, denoting $J_{0}$ and $J_{1}$ as the zeroth and first order Bessel functions of the first kind, respectively.}\label{fig3}
\end{figure}
    \section{C-PHASE GATE OF TWO RYDBERG ATOMS}\label{sec3}
    In this section, we utilize the FFM scheme to construct a universal two-qubit C-Phase gate, capitalizing on its intricate dynamics. By varying the modulation index $\alpha$ and modulation frequency $\omega_0$, we present the time-average population~\cite{PhysRevLett.120.123204,Zhao2023} of the doubly excited Rydberg state $|rr\rangle$ within 10 $\mu$s in Fig.~\ref{fig2}(a) with the RRI strength of $V=20\Omega$. The results indicate that the FFM scheme can effectively manipulate the dynamic behavior of system according to the population behaviors of $|rr\rangle$. Notably, when the resonance condition $\omega_0=V$ is satisfied, the population of the $|rr\rangle$ state is primarily modulated by $J_{0}(\alpha)$ and $J_{1}(\alpha)$. By selecting values of $\alpha$ at which either $J_{0}(\alpha)$ or $J_{1}(\alpha)$ vanishes, the population of $|rr\rangle$ state can be effectively suppressed.

	\begin{figure*}
	\centering
	\includegraphics[width=18cm,height=5.52cm]{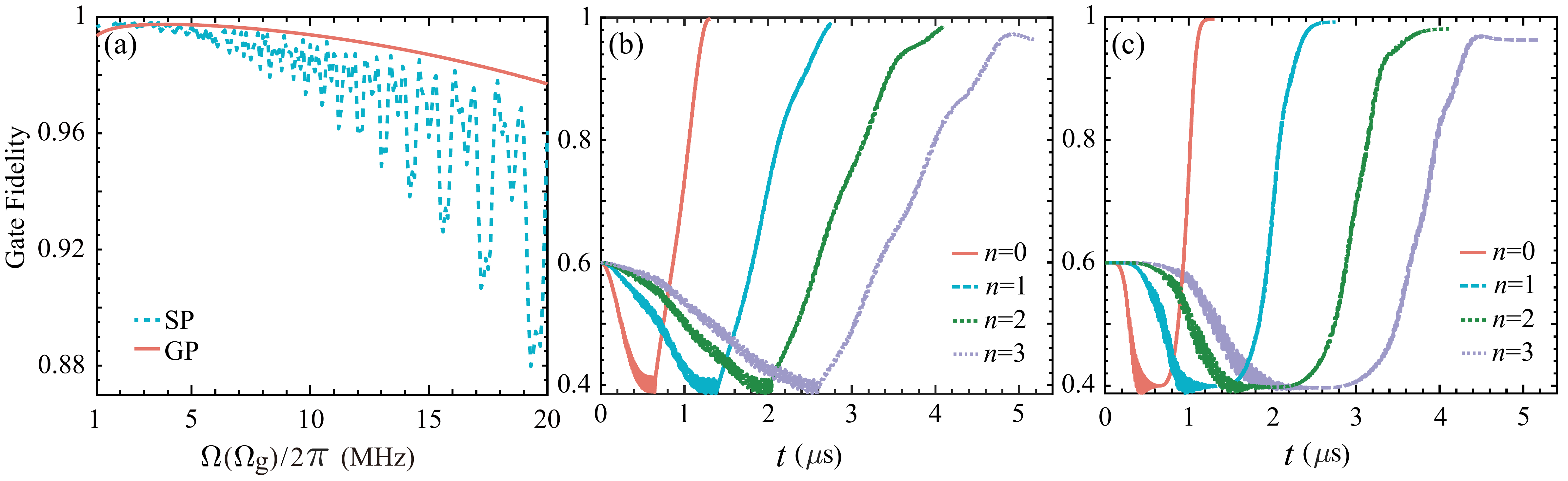}
	\caption{(a) The average fidelity of the constructed C-Phase gate fluctuates as the square pulse Rabi frequency $\Omega$ and the Gaussian pulse maximum amplitude $\Omega_g$ ranges from $2\pi\times1$ MHz to $2\pi\times20$ MHz when $n$ = 0. (b) The evolution of the average gate fidelity of the implemented C-Phase gate over time, when $n$ equals 0,1,2,3. The time-independent Rabi frequency $\Omega=2\pi\times3.5$ MHz, V = $\omega_0$ = $2\pi\times70.18$ MHz, and the gate duration being $T = 2\pi/|\Omega_a(t)|$. (c) The temporal progression of the average gate fidelity for the C-Phase gate utilizing Gaussian soft quantum control when $n$ = 0,1,2,3. The time-dependent Rabi frequency is given by $\Omega_g(t)$ with the maximum amplitude $\Omega_g=2\pi\times8.1$ MHz , $T_g = (-1+a)\pi/|\Omega_g(t)|J_{0}(\alpha)(4a-\sqrt{\pi})$ and a gate time of $T = 8T_g$. Additionally, $V = \omega_0=2\pi\times70.18$ MHz.}\label{fig4}
\end{figure*}
	To construct the desired two-qubit C-Phase gate $\hat{U}_{CP} =$ diag $(1,-e^{i\vartheta},-e^{i\vartheta},1)$, we employ two sequential identical Rydberg global pulses with equal duration $\tau = \pi/|\Omega_a(t)|$ but a laser phase jump $\vartheta$ inserted between the two pulses. Utilizing the unitary operator $\hat{U}_f=\operatorname{exp}(-i\int_0^\tau\hat{H}_i(t)dt)$, where $i$ = 01, 10, 11, each pulse induces a transformation in the atomic states. An in-depth exploration into the evolution of the four computational basis states under the impact of $\hat{U}_{CP}$ is presented in Fig.~\ref{fig2}(b). The state $|00\rangle$ is unaffected by these pulses with $\hat{U}_f|00\rangle = |00\rangle$, while the states $|01\rangle$ and $|10\rangle$ return to themselves with an additional phase factor represented by $\hat{U}_f|01\rangle(|10\rangle) = -e^{i\vartheta}|01\rangle(|10\rangle)$ after a duration of $2\tau$. Lastly, concerning the evolution of the state $|11\rangle$, our objective is to return it to the same state without accumulating any phase after the complete evolution. This is achieved by appropriately selecting the time-dependent functions $|\Omega_a(t)|$ and $|\Omega_b(t)|$.
    Thus, we define $|\Omega_b(t)|$ as $N$ times $|\Omega_a(t)|$ and evaluate the time evolution operator $\hat{U}_f$ for the corresponding three-level system using Eq.~\eqref{eq6}. By applying the condition $\hat{U}_f|11\rangle = |11\rangle$, we can analyze the changes in population $P_{11}$ and phase $\Phi_{11}$ of the state $|11\rangle$ as $N$ varies. It can be seen that when 
    \begin{equation}\label{eq7}
    	N = (\sqrt{31}-\sqrt{7})n+\sqrt{7}, ~n=0,1,2,\cdots
    \end{equation}
    the phase $\Phi_{11}=0$ is satisfied, the numerical result shown in Fig.~\ref{fig3}(a) also prove this, and the state $|11\rangle$ undergoes self-evolution without accumulating any additional phase. In other words, we can choose various values of $n$ for the realization of the universal C-Phase gate, which undoubtedly demonstrates the unique advantages of our FFM scheme, enriching the dynamics of the quantum system and offering diversified options for the construction of quantum gates. With increasing $n$, the population of the state $|11\rangle$ remains close to 1 due to the significant disparity between $|\Omega_b(t)|$ and $|\Omega_a(t)|$. This intriguing behavior, similar to the unconventional Rydberg pumping mechanism~\cite{Li2018Unconventional}, effectively freezes the evolution of a two-atom system featuring the same ground states.

    To satisfy the resonance conditions detailed in Eq.~\eqref{eq4}, we assign $\omega_{0}=V$ such that the coupling strength between the singly-excited state $|W\rangle$ and the doubly-excited state  $|rr\rangle$ is primarily influenced by $J_{1}(\alpha)$, while the Rabi frequency between the state $|11\rangle$ and state $|W\rangle$ is predominantly governed by $J_{0}(\alpha)$. This configuration ensures that $|\Omega_b(t)|=N|\Omega_a(t)|$ is equivalent to $J_{1}(\alpha)=NJ_{0}(\alpha)$, enabling us to select a suitable modulation index $\alpha$ to attain the desired nontrivial two-qubit gate illustrated in Fig.~\ref{fig3}(b). The aforementioned FFM approach facilitates the realization of the two-qubit C-Phase gate: $\hat{U}_{CP} =$ diag $(1,-e^{i\vartheta},-e^{i\vartheta},1)$ in reference to the computational basis $\{|00\rangle,|01\rangle,|10\rangle,|11\rangle\}$.
	
		\begin{figure*}
		\centering
		\includegraphics[width=18cm,height=5.42cm]{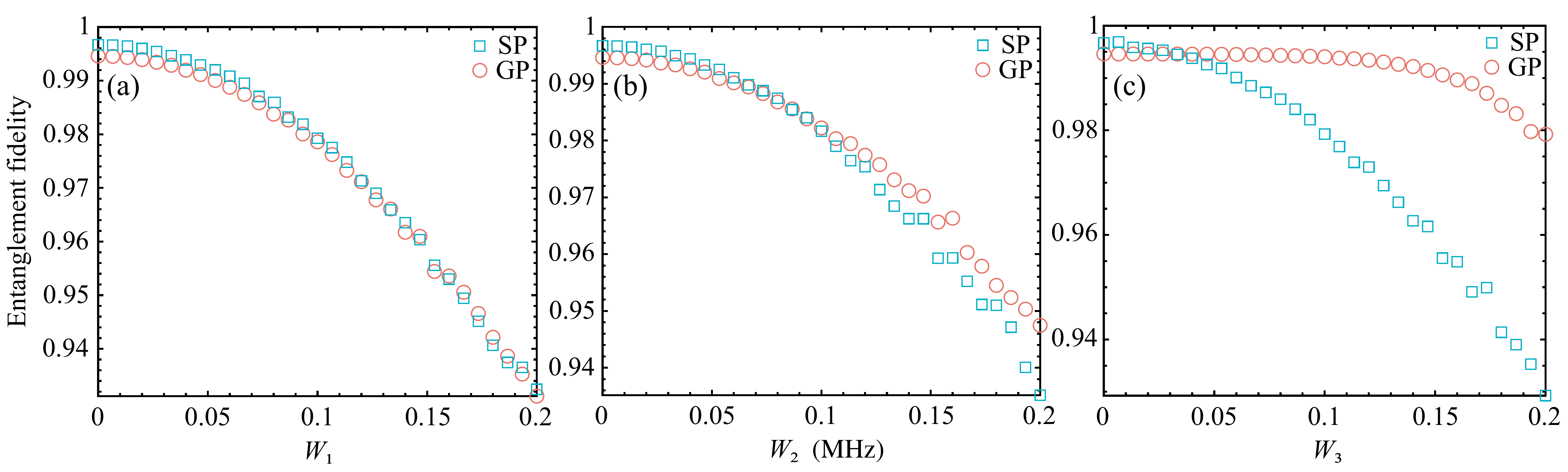}
		\caption{The impact of errors in the (a) Rabi frequency $\Omega$, (b) detuning $\Delta_0$ and (c) gate time $T$ on state fidelity is examined using square pulse and Gaussian pulse configurations. The relevant parameters remain consistent with $V = \omega_0=2\pi\times70.18$ MHz, $\Omega=2\pi\times3.5$ MHz, and the time-dependent Rabi frequency $\Omega_g(t)$ as previously described. Each point denotes the average of 1001 results obtained by randomly picking 1001 disorders from $[-W_{k},W_{k}]$ with $k = 1, 2$, and $3$ in (a), (b) and (c), respectively.}\label{fig5}
	\end{figure*}
	\section{Fidelity and robustness}\label{sec4}
	
	\subsection{Parameter selection and numerical simulation}\label{sec4.1}
	
	In this section, we employ feasible experimental parameters to conduct a numerical simulation for testing the performance of our FFM scheme. 	The excitation process from the ground state $|1\rangle$ to the Rydberg state $|r\rangle$ can be achieved via a two-photon mechanism in $^{87}$Rb atoms~\cite{PhysRevLett.123.170503,PhysRevLett.121.123603,Evered2023}. Within the FFM scheme, to meet the condition $V=\omega_0\gg\Omega$, it is necessary to have Rydberg states characterized by sufficiently high principal quantum numbers or close interatomic separations to ensure a potent RRI strength. Hence, we consider energy levels as two hyperfine ground states $|0\rangle = |5S_{1/2}, F = 1, m_F = 1\rangle$ and $|1\rangle = |5S_{1/2}, F = 2, m_F = 2\rangle$~\cite{PhysRevLett.104.010502}, an intermediary state $|p\rangle = |5p_{3/2}\rangle$ or $|p\rangle = |6p_{3/2}\rangle$, and the Rydberg strongly-interacting state $|r\rangle = |70S_{1/2}\rangle$ possessing an interaction coefficient of $C_6/2\pi = 858.4$~GHz $\mu \rm m^6$. For an accessible atomic temperature $T_a = 10~{\rm \mu K}$, the lifetime $\tau_r$ of the Rydberg state $|r\rangle$ in $^{87}$Rb atoms with a principal quantum number $70$ is approximately $400~{\rm \mu s}$~\cite{PhysRevLett.123.170503}. The strength of RRI is set as $V = 2\pi\times70.18$~MHz, considering an interatomic separation of $d=4.8~{\rm \mu m}$~\cite{Bernien2017,PhysRevLett.121.123603}. 
	
	To study evolution of the two-atom system, we utilize the Lindblad master equation to assess the gate performance under dissipative effects, expressed as
	\begin{equation}\label{eq8}
		\frac{\partial\hat\rho(t)}{\partial t}=-\frac{i}{\hbar}[\hat{H}(t)_{},\hat\rho(t)]+\mathcal{L}_{r}[\hat\rho]
	\end{equation}
    where $\hat\rho(t)$ represents the density matrix operator corresponding to the two-atom system, $\hat{H}(t)$ stands for the time-varying Hamiltonian of the system as depicted in Eq.~\eqref{eq1}, and the dissipation terms $\mathcal{L}_{r}[\hat\rho]$ account for the spontaneous decay originating from the strongly-interacting states, elucidated as
	\begin{equation}
	\mathcal{L}_{r}[\hat\rho]=\sum_{i=1,2}^{}\sum_{j=0,1}^{}(\hat{L}_{j}^i\hat\rho \hat{L}_{j}^{i\dagger}-\frac12\hat{L}_{j}^{i\dagger}\hat{L}_{j}^i\hat\rho-\frac12\hat\rho\hat{L}_{j}^{i\dagger}\hat{L}_{j}^i)
	\end{equation}
	in which the Lindblad operators $\hat{L}_{j}^i = \sqrt{\gamma_j}|j\rangle_{i}\langle r|$ represent the jump operators characterizing the spontaneous emission of the atom from the Rydberg state $|r\rangle$ to a ground state $|j\rangle$, where $\gamma_j$ represents the decay rate.
	
	To demonstrate the efficacy and resilience of our proposed two-qubit C-Phase gate across diverse initial conditions, we evaluate the average fidelity~\cite{NIELSEN2002249}
	\begin{equation}
		\bar{F}(\xi,\hat{U})=\frac{\sum_k\operatorname{tr}[\hat{U}\hat{U}_k^\dagger\hat{U}^\dagger\xi(\hat{U}_k)]+d_k^2}{d_k^2(d_k+1)},
	\end{equation}
	in which $\hat{U}$ represents the perfect logic gate, $\hat{U}_k$ stands for the tensor of Pauli matrices $\hat{I}\hat{I},\hat{I}\hat{\sigma}_x,\hat{I}\hat{\sigma}_y,\hat{I}\hat{\sigma}_z,...,\hat{\sigma}_z\hat{\sigma}_z$ for a two-qubit gate, $d_k = 2^x$, with $x$ being the number of qubits in a quantum gate and in this case, $d_k = 4$. $\xi(\hat{U}_k)$ denotes the trace-preserving quantum operation achieved by solving the master equation \eqref{eq8}.

	As illustrated by the cyan dashed line in Fig.~\ref{fig4}(a), the average fidelity of the C-Phase gate we have constructed varies with the Rabi frequency ranging from $2\pi\times1$ MHz to $2\pi\times20$ MHz when $n = 0$, which indicates that the Rydberg blockade condition $\Omega\ll V$ should be satisfied to attain a relatively higher and stable gate fidelity. For subsequent analysis, we selected a time-independent Rabi frequency of $2\pi\times3.5$~MHz. In Fig.~\ref{fig4}(b), we present the evolving average fidelity of the C-Phase gate with $\vartheta=\pi/2$ for $n = 0, 1, 2, 3$ corresponding to the gate time $T = 2\pi/|\Omega_a(t)|$, yielding final average fidelities of 0.9973, 0.9870, 0.9856 and 0.9647, respectively. It can be observed that for $n = 0$, the achieved gate fidelity is the highest with the shortest gate time. Conversely, for $n = 1, 2$ and $3$, the gate time increases due to the decreasing values of selected $J_{0}(\alpha)$ and $J_{1}(\alpha)$, leading to a decline in the gate fidelity influenced by the non-resonant terms and dissipation of system.

	\subsection{Gaussian soft control}\label{sec4.2}
		Observing from Fig.~\ref{fig4}(b) and the cyan dashed line of Fig.~\ref{fig4}(a), we note that the fidelity evolution exhibits pronounced oscillations due to imperfect parameter settings aimed at mitigating the high-frequency oscillations, following the defective RWA. These substantial oscillations render the resultant C-Phase gate particularly susceptible to control errors. To address this issue, we leverage the technique of soft quantum control ~\cite{PhysRevLett.121.050402} to refine the parameters for better adherence to RWA conditions, thus damping the high-frequency oscillatory effects and reducing parameter uncertainties during the phase accumulation, ultimately enhancing the gate performance.
	
    Under the implementation of soft quantum control, the time-independent square pulse~(SP) with Rabi frequency $\Omega$ can be transformed into a time-evolving Gaussian pulse~(GP), given by
    \begin{equation}
   	\Omega_{g}(t)=\begin{cases}\Omega_{g}\left[e^{\frac{-(t-2T_{g})^2}{T_{g}^2}}-a\right]/(1-a), & 0<t\le4T_{g},\\  & \\ \Omega_{g}\left[e^{\frac{-(t-6T_{g})^2}{T_{g}^2}}-b\right]/(1-b), & 4T_{g}<t\le T,\end{cases}
    \end{equation}
    where $\Omega_g$ and $T_g$ denote the maximum amplitude and width of GP, respectively. Furthermore, $a =$ exp $[-(2T_g)^2/T_g^2]$ and $b =$ exp $[-(T/2-6T_g)^2/T_g^2]$ causing the amplitude is zero at the start and end. Ensuring the pulse area remains constant, we deduce  $T_g = (-1+a)\pi/|\Omega_g(t)|J_{0}(\alpha)(4a-\sqrt{\pi})$ by satisfying $\int_0^{4T_{g}}\Omega_g(t)dt=\pi$ and $\int_{4T_{g}}^{t_{g}}\Omega_g(t)dt=\pi$ with the gate time $T= 8T_g$. To highlight the distinctive superiority of the GP over the SP, the variation in gate fidelity with respect to the GP's maximum amplitude under soft control is illustrated by the solid red line in Fig.~\ref{fig4}(a). The findings demonstrate that this approach noticeably enhances gate performance and facilitates a more gradual evolution of fidelity. Employing this GP configuration, we choose the maximum amplitude $\Omega_g=2\pi\times8.1$ MHz so that the optimized gate time is consistent with the square pulse and illustrate the average gate fidelity of the C-Phase gate for the cases $n = 0,1,2,3$ in Fig.~\ref{fig4}(c). The results depict a substantial reduction in oscillations through Gaussian soft quantum control. Moreover, GP offers practical advantages over traditional SP in experimental scenarios~\cite{Evered2023}.  Consequently, the combination of our approach with Gaussian soft control significantly enhances operational feasibility in experiments, showcasing promising and valuable applications.
	
			\begin{figure}[t]
		\centering
		\includegraphics[width=0.9\linewidth]{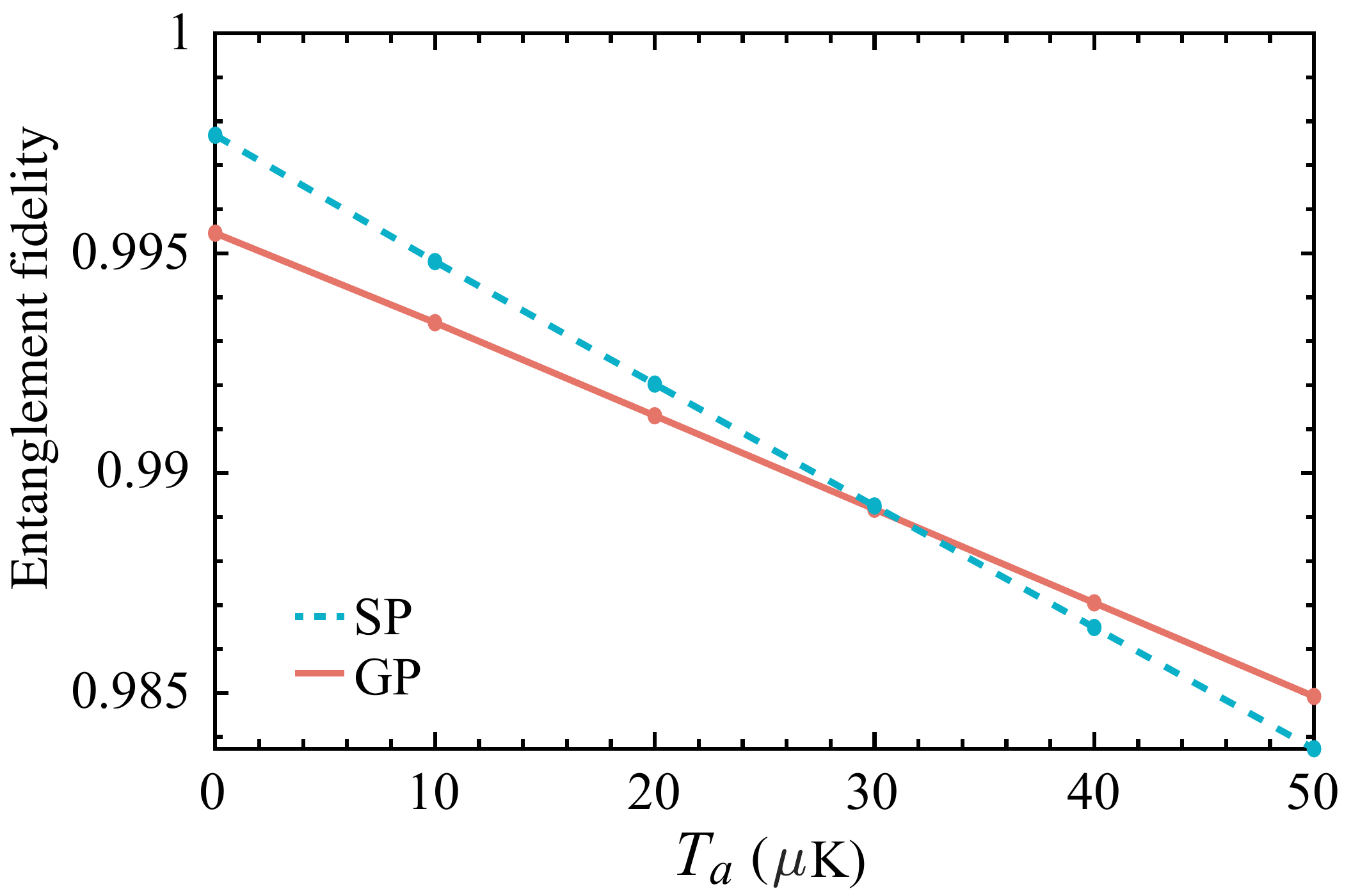}
		\caption{Impact of Doppler dephasing errors on state fidelity at various temperatures for square pulse and Gaussian pulse. The associated parameters remain consistent with the aforementioned ones.}\label{fig6}
	\end{figure}
	\subsection{Gate resilience}\label{sec4.3}	
	Subsequently, we delve into the robustness of the constructed C-Phase gates against parameter inaccuracies, with a particular focus on the impact of laser intensity on the fidelity of the C-Phase gate. To explore this scenario, we examine the effects of deviations in the Rabi frequency $\Omega$, detuning $\Delta_0$ and gate time $T$, encompassing both upward and downward deviations. To quantify these deviations, we introduce the deviation value $\delta\Omega$, $\delta\Delta_0$ and $\delta T$, yielding actual parameters
	    \begin{equation}
	\Omega\rightarrow\Omega(1+W_1),\quad\Delta_0\rightarrow\Delta_0+W_2,\quad T \rightarrow T(1+W_3).
	\end{equation}
Subsequently, we analyze the influence of errors on the entanglement fidelity of an equivalent Bell state $|\Psi_{\rm Bell}\rangle=(|00\rangle-i|01\rangle-i|10\rangle+|11\rangle)/2$. The entanglement fidelity is defined by ${\rm tr}[{\hat \rho}(T)|\Psi_{\rm Bell}\rangle\langle\Psi_{\rm Bell}|]$, where density operator ${\hat \rho}(T)$ at the final moment is obtained by solving the master equation with an initial two-atom product state $|\psi_0\rangle=(|0\rangle+|1\rangle)\otimes(|0\rangle+|1\rangle)/2$ when employing either SP or GP for the laser configuration, as illustrated in Figs.~\ref{fig5}(a)-(c). We observe that the entanglement fidelity exhibits a comparable level of performance degradation when subjected to the same relative error in either pulse strength or detuning parameters, regardless of whether a SP pulse or GP is employed. This phenomenon can be attributed to the fundamental dependence of quantum evolution on the effective pulse area. Specifically, an equivalent relative error in the pulse strength parameter across both scenarios results in identical alterations to the effective pulse area. Consequently, precise control over pulse strength and detuning emerges as a critical requirement for achieving high-fidelity C-Phase gate operations within the present gate implementation scheme. Alternatively, it is noteworthy that the robustness of C-Phase gates against pulse strength errors can be significantly enhanced through optimal control methodologies. These techniques introduce additional degrees of freedom into the Rydberg system~\cite{PhysRevA.109.022613,PhysRevApplied.21.064053,PhysRevA.103.062607,PhysRevA.105.012611}, thereby potentially improving error tolerance. However, this enhanced robustness comes at the cost of increased implementation complexity, which may inadvertently introduce new sources of error into the system. Regarding the timing error in Fig.~\ref{fig5}(c), the entanglement fidelity demonstrates remarkable resilience when the GP is used. This robustness stems from the inherent property that moderate deviations in operation time do not significantly alter the pulse area. Only when the actual pulse duration falls substantially below the desired value, does the entanglement fidelity exhibit a marginal decrease. This behavior stands in stark contrast to the SP gate implementation, where the entanglement fidelity shows pronounced sensitivity to timing errors. Such sensitivity arises from the direct linear relationship between pulse duration errors and pulse area variations when the SP is used.

	Finally, we simulate the dynamics of quantum systems with Doppler dephasing errors~\cite{SIBALIC2017319,PhysRevApplied.13.024008}. The Rydberg atom, regardless of its temperature, is not at rest in the trap. Doppler dephasing caused by the random motion of imprisoned atoms is one of the most important sources of error in experiments, which will limit the performance of quantum gates. Due to the presence of the Doppler effect, the Rydberg excitation Rabi frequency changes to $\Omega(t)\to\Omega(t)e^{i\delta t}$, where $\delta=k_{\mathrm{eff}}\Delta v$ denotes the Doppler shift with effective laser wavevector $k_{\mathrm{eff}}$ and atomic root-mean-square speed $\Delta v =\sqrt{k_{\mathrm{B}}T_{a}/m}$. In which $k_{\mathrm{B}}$, $T_a$, and $m$ represent the Boltzmann constant, atomic temperature, and atomic mass, respectively. For the two-photon transition process of Rb atoms, we choose laser wavelengths of $\lambda_{1}$ = 780 nm and $\lambda_{2}$ = 480 nm, respectively. Moreover, we replace the orthogonal lasers with the counterpropagating lasers to reduce the effect of Doppler dephasing, thus obtaining $k_{\mathrm{eff}} = 2\pi(1/\lambda_{2}-1/\lambda_{1})$. The numerical results of the fidelity of the C-Phase gate we constructed are shown in Fig.~\ref{fig6}, and it should be noted that the temperature simulated here is much higher than the current experimental conditions, where the temperature of the atoms in the optical tweezers can be cooled to 5.2 $\mu$K~\cite{PhysRevA.99.043404}. Even so, in the presence of a Doppler shift, the fidelity of the quantum gate can still reach more than 0.98 for cases of  the SP and the GP. Besides, the usage of GP exhibits better performance in resilience against the Doppler dephasing, knowing from a smaller slope of the red entanglement fidelity line in Fig.~\ref{fig6}.

		\begin{figure}[t]
		\centering
		\includegraphics[width=\linewidth]{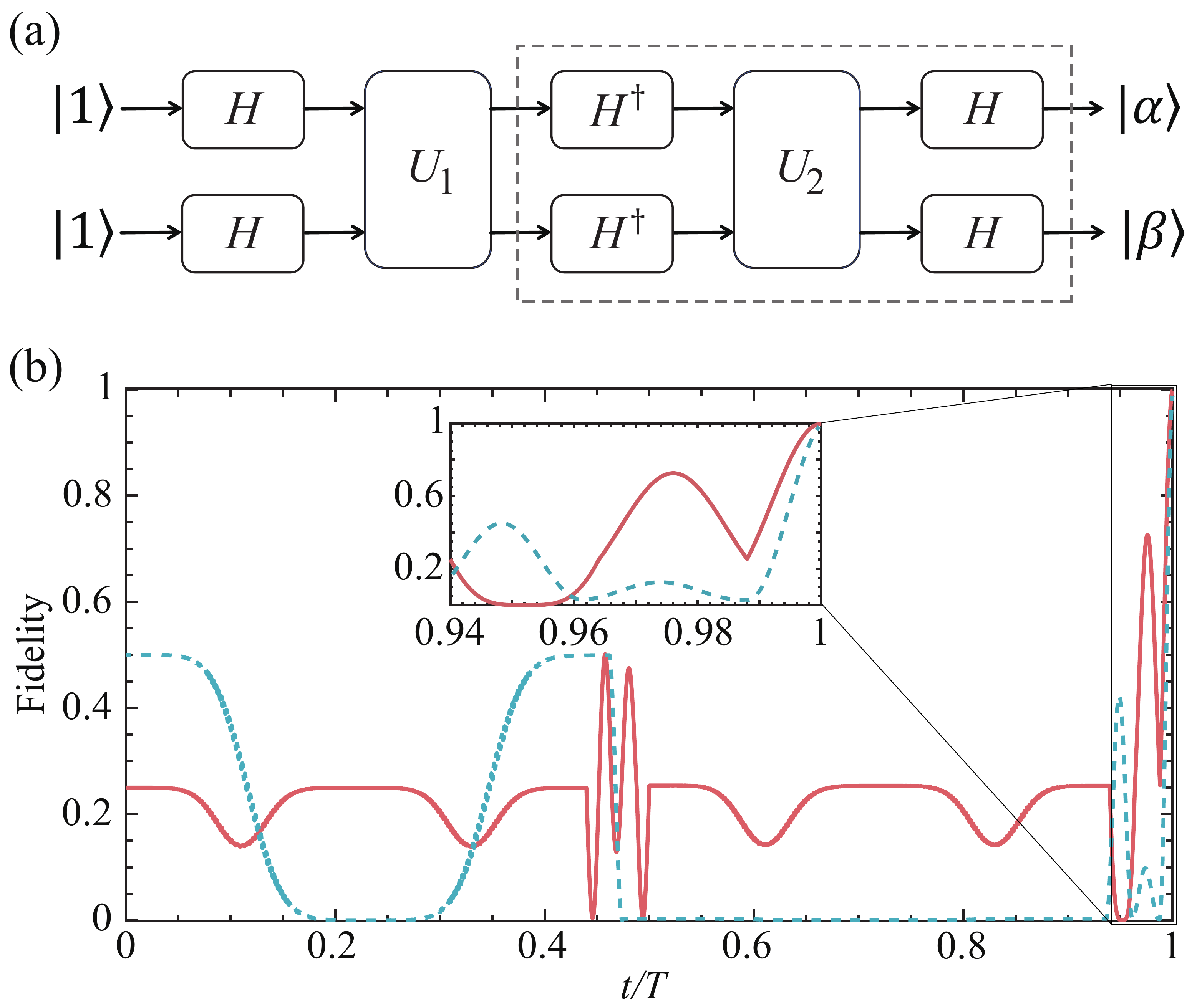}
		\caption{(a) The quantum circuit depicting the search algorithm for both one-item and two-item scenarios. (b) The fidelity evolution of the target state in the search algorithm for one-item and two-item over time, where the solid red line denotes the one-item and the dashed blue line represents the two-item.}\label{fig7}
	\end{figure}
	\section{Application in the Grover-Long algorithm}\label{sec5}
	
	Finally, we consider to apply the proposed C-Phase gates in implementing the multi-item quantum search algorithm. According to the Grover-Long algorithm~\cite{LONG2002143,PhysRevA.106.052610}, by replacing the phase inversion with phase rotation in oracle- and diffusion-operation, the desired target items can be searched with zero failure rate. The detailed circuit diagram is presented in Fig.~\ref{fig7}(a), together with the following steps.
	
	(i)~Starting with the initial state $\frac12(|00\rangle+|01\rangle+|10\rangle+|11\rangle)$, the oracle operator is applied to mark the target items. The utilization of one-item search $|11\rangle$ as a case study to elucidate the implementation procedure, the oracle operator is defined as $U_1$ = $|00\rangle\langle 00|+|01\rangle\langle 01|+|10\rangle\langle 10|-|11\rangle\langle 11|$, which is implemented by selecting $\vartheta=\pi/2$ in C-Phase gate together with the single-qubit operator $U$ = $|0\rangle\langle 0|+i|1\rangle\langle 1|$ applied to both qubits. For the two-item target state $(|01\rangle+|10\rangle)/\sqrt2$, the oracle operator is given by $U_1$ = $|00\rangle\langle 00|+i|01\rangle\langle 01|+i|10\rangle\langle 10|+|11\rangle\langle 11|$, which is realized by setting $\vartheta=-\pi/2$ in C-Phase gate.

    (ii)~For the diffusion operator, the inverse Hadamard operation ($H^{\dag}$) is initially applied to each qubit. The operator $U_2$ is identical to $U_1$ in the case of the one-item search. However, for the two-item search, $U_2$ is defined as $|00\rangle\langle 00|+|01\rangle\langle 01|+|10\rangle\langle 10|+i|11\rangle\langle 11|$, which is implemented by setting $\vartheta=-5\pi/4$ in C-Phase gate together with the single qubit operation $U$ = $|0\rangle\langle 0|+e^{i\pi/4}|1\rangle\langle 1|$ applied to each qubit. Finally, the Hadamard operation is applied to each qubit to complete the diffusion process.

	Following these steps, the one-item and two-item searches are successfully implemented, with the dynamical processes are illustrated in Figs.~\ref{fig7}(b). The achieved fidelities of 0.9975 for the one-item search and 0.9879 for the two-item search underscore the effectiveness of our scheme. Notably, the proposed approach simplifies the circuit design and eliminates the need for individual addressing in two-qubit operations, thereby significantly improving the accuracy and efficiency of the quantum search algorithm.

	\section{CONCLUSION}\label{sec6}
	
	We have introduced a strategy to realize a C-Phase gate utilizing Rydberg atoms with FFM techniques. This approach is implemented through the Floquet periodic modulation of the atom-field detuning. This method leverages the synergistic effect between periodic modulation and RRI strength, providing a versatile platform for quantum system manipulation through parameter optimization. Notably, it effectively circumvents the conventional limitation where interaction strength is strictly governed by interatomic distance in Rydberg systems. This capability is particularly advantageous for the development of quantum logic gates. With respect to the most recent experimental parameters concerning the alkali metal rubidium atom, the C-Phase gates devised by our proposed approach exhibit outstanding performances. Furthermore, our method can be integrated with the Gaussian soft quantum control to enhance the fidelity and robustness of the C-Phase gate. Ultimately, our C-Phase holds promise for search algorithms with high success rates, thereby significantly enhancing the efficacy of multi-qubit operations without the necessity for laser independent addressing. We envision that combining FFM with advanced quantum control of Rydberg atoms will unlock new opportunities for achieving efficient and fault-tolerant quantum computation, enabling the realization of sophisticated quantum algorithms and the discovery of unprecedented quantum phenomena in precisely engineered coherent systems.

\section*{acknowledgements}
The authors acknowledge the financial support by the National Natural Science Foundation of China~(Grants No.~12304407, No.~12274376, No. 62471001, No. 12475009, No. 12075001, and No. 12175001), the Major Science and Technology project of Henan Province under Grant No. 221100210400, the Natural Science Foundation of Henan Province under Grant No. 232300421075, Natural Science Research Project in Universities of Anhui Province (No. 2024AH050068), Anhui Provincial Key Research and Development Plan (Grant No. 2022b13020004), Anhui Province Science and Technology Innovation Project (Grant No. 202423r06050004), and the China Postdoctoral Science Foundation~(Grants No.~2023TQ0310, No.~GZC20232446, and No.~2024M762973).

		\bibliography{REV}

\begin{thebibliography}{92}%
\makeatletter
\providecommand \@ifxundefined [1]{%
 \@ifx{#1\undefined}
}%
\providecommand \@ifnum [1]{%
 \ifnum #1\expandafter \@firstoftwo
 \else \expandafter \@secondoftwo
 \fi
}%
\providecommand \@ifx [1]{%
 \ifx #1\expandafter \@firstoftwo
 \else \expandafter \@secondoftwo
 \fi
}%
\providecommand \natexlab [1]{#1}%
\providecommand \enquote  [1]{``#1''}%
\providecommand \bibnamefont  [1]{#1}%
\providecommand \bibfnamefont [1]{#1}%
\providecommand \citenamefont [1]{#1}%
\providecommand \href@noop [0]{\@secondoftwo}%
\providecommand \href [0]{\begingroup \@sanitize@url \@href}%
\providecommand \@href[1]{\@@startlink{#1}\@@href}%
\providecommand \@@href[1]{\endgroup#1\@@endlink}%
\providecommand \@sanitize@url [0]{\catcode `\\12\catcode `\$12\catcode
  `\&12\catcode `\#12\catcode `\^12\catcode `\_12\catcode `\%12\relax}%
\providecommand \@@startlink[1]{}%
\providecommand \@@endlink[0]{}%
\providecommand \url  [0]{\begingroup\@sanitize@url \@url }%
\providecommand \@url [1]{\endgroup\@href {#1}{\urlprefix }}%
\providecommand \urlprefix  [0]{URL }%
\providecommand \Eprint [0]{\href }%
\providecommand \doibase [0]{http://dx.doi.org/}%
\providecommand \selectlanguage [0]{\@gobble}%
\providecommand \bibinfo  [0]{\@secondoftwo}%
\providecommand \bibfield  [0]{\@secondoftwo}%
\providecommand \translation [1]{[#1]}%
\providecommand \BibitemOpen [0]{}%
\providecommand \bibitemStop [0]{}%
\providecommand \bibitemNoStop [0]{.\EOS\space}%
\providecommand \EOS [0]{\spacefactor3000\relax}%
\providecommand \BibitemShut  [1]{\csname bibitem#1\endcsname}%
\let\auto@bib@innerbib\@empty
\bibitem [{\citenamefont {Jaksch}\ \emph {et~al.}(2000)\citenamefont {Jaksch},
  \citenamefont {Cirac}, \citenamefont {Zoller}, \citenamefont {Rolston},
  \citenamefont {C\^ot\'e},\ and\ \citenamefont {Lukin}}]{PhysRevLett.85.2208}%
  \BibitemOpen
  \bibfield  {author} {\bibinfo {author} {\bibfnamefont {D.}~\bibnamefont
  {Jaksch}}, \bibinfo {author} {\bibfnamefont {J.~I.}\ \bibnamefont {Cirac}},
  \bibinfo {author} {\bibfnamefont {P.}~\bibnamefont {Zoller}}, \bibinfo
  {author} {\bibfnamefont {S.~L.}\ \bibnamefont {Rolston}}, \bibinfo {author}
  {\bibfnamefont {R.}~\bibnamefont {C\^ot\'e}}, \ and\ \bibinfo {author}
  {\bibfnamefont {M.~D.}\ \bibnamefont {Lukin}},\ }\bibfield  {title} {\enquote
  {\bibinfo {title} {Fast quantum gates for neutral atoms},}\ }\href {\doibase
  10.1103/PhysRevLett.85.2208} {\bibfield  {journal} {\bibinfo  {journal}
  {Phys. Rev. Lett.}\ }\textbf {\bibinfo {volume} {85}},\ \bibinfo {pages}
  {2208--2211} (\bibinfo {year} {2000})}\BibitemShut {NoStop}%
\bibitem [{\citenamefont {Gallagher}(1994)}]{Gallagher_1994}%
  \BibitemOpen
  \bibfield  {author} {\bibinfo {author} {\bibfnamefont {Thomas~F.}\
  \bibnamefont {Gallagher}},\ }\href@noop {} {\emph {\bibinfo {title} {Rydberg
  Atoms}}},\ Cambridge Monographs on Atomic, Molecular and Chemical Physics\
  (\bibinfo  {publisher} {Cambridge University Press},\ \bibinfo {year}
  {1994})\BibitemShut {NoStop}%
\bibitem [{\citenamefont {Saffman}\ \emph {et~al.}(2010)\citenamefont
  {Saffman}, \citenamefont {Walker},\ and\ \citenamefont
  {M\o{}lmer}}]{RevModPhys.82.2313}%
  \BibitemOpen
  \bibfield  {author} {\bibinfo {author} {\bibfnamefont {M.}~\bibnamefont
  {Saffman}}, \bibinfo {author} {\bibfnamefont {T.~G.}\ \bibnamefont {Walker}},
  \ and\ \bibinfo {author} {\bibfnamefont {K.}~\bibnamefont {M\o{}lmer}},\
  }\bibfield  {title} {\enquote {\bibinfo {title} {Quantum information with
  rydberg atoms},}\ }\href {\doibase 10.1103/RevModPhys.82.2313} {\bibfield
  {journal} {\bibinfo  {journal} {Rev. Mod. Phys.}\ }\textbf {\bibinfo {volume}
  {82}},\ \bibinfo {pages} {2313--2363} (\bibinfo {year} {2010})}\BibitemShut
  {NoStop}%
\bibitem [{\citenamefont {B\'eguin}\ \emph {et~al.}(2013)\citenamefont
  {B\'eguin}, \citenamefont {Vernier}, \citenamefont {Chicireanu},
  \citenamefont {Lahaye},\ and\ \citenamefont
  {Browaeys}}]{PhysRevLett.110.263201}%
  \BibitemOpen
  \bibfield  {author} {\bibinfo {author} {\bibfnamefont {L.}~\bibnamefont
  {B\'eguin}}, \bibinfo {author} {\bibfnamefont {A.}~\bibnamefont {Vernier}},
  \bibinfo {author} {\bibfnamefont {R.}~\bibnamefont {Chicireanu}}, \bibinfo
  {author} {\bibfnamefont {T.}~\bibnamefont {Lahaye}}, \ and\ \bibinfo {author}
  {\bibfnamefont {A.}~\bibnamefont {Browaeys}},\ }\bibfield  {title} {\enquote
  {\bibinfo {title} {Direct measurement of the van der waals interaction
  between two rydberg atoms},}\ }\href {\doibase
  10.1103/PhysRevLett.110.263201} {\bibfield  {journal} {\bibinfo  {journal}
  {Phys. Rev. Lett.}\ }\textbf {\bibinfo {volume} {110}},\ \bibinfo {pages}
  {263201} (\bibinfo {year} {2013})}\BibitemShut {NoStop}%
\bibitem [{\citenamefont {Omran}\ \emph {et~al.}(2019)\citenamefont {Omran},
  \citenamefont {Levine}, \citenamefont {Keesling}, \citenamefont {Semeghini},
  \citenamefont {Wang}, \citenamefont {Ebadi}, \citenamefont {Bernien},
  \citenamefont {Zibrov}, \citenamefont {Pichler}, \citenamefont {Choi},
  \citenamefont {Cui}, \citenamefont {Rossignolo}, \citenamefont {Rembold},
  \citenamefont {Montangero}, \citenamefont {Calarco}, \citenamefont {Endres},
  \citenamefont {Greiner}, \citenamefont {Vuletić},\ and\ \citenamefont
  {Lukin}}]{doi:10.1126/science.aax9743}%
  \BibitemOpen
  \bibfield  {author} {\bibinfo {author} {\bibfnamefont {A.}~\bibnamefont
  {Omran}}, \bibinfo {author} {\bibfnamefont {H.}~\bibnamefont {Levine}},
  \bibinfo {author} {\bibfnamefont {A.}~\bibnamefont {Keesling}}, \bibinfo
  {author} {\bibfnamefont {G.}~\bibnamefont {Semeghini}}, \bibinfo {author}
  {\bibfnamefont {T.~T.}\ \bibnamefont {Wang}}, \bibinfo {author}
  {\bibfnamefont {S.}~\bibnamefont {Ebadi}}, \bibinfo {author} {\bibfnamefont
  {H.}~\bibnamefont {Bernien}}, \bibinfo {author} {\bibfnamefont {A.~S.}\
  \bibnamefont {Zibrov}}, \bibinfo {author} {\bibfnamefont {H.}~\bibnamefont
  {Pichler}}, \bibinfo {author} {\bibfnamefont {S.}~\bibnamefont {Choi}},
  \bibinfo {author} {\bibfnamefont {J.}~\bibnamefont {Cui}}, \bibinfo {author}
  {\bibfnamefont {M.}~\bibnamefont {Rossignolo}}, \bibinfo {author}
  {\bibfnamefont {P.}~\bibnamefont {Rembold}}, \bibinfo {author} {\bibfnamefont
  {S.}~\bibnamefont {Montangero}}, \bibinfo {author} {\bibfnamefont
  {T.}~\bibnamefont {Calarco}}, \bibinfo {author} {\bibfnamefont
  {M.}~\bibnamefont {Endres}}, \bibinfo {author} {\bibfnamefont
  {M.}~\bibnamefont {Greiner}}, \bibinfo {author} {\bibfnamefont
  {V.}~\bibnamefont {Vuletić}}, \ and\ \bibinfo {author} {\bibfnamefont
  {M.~D.}\ \bibnamefont {Lukin}},\ }\bibfield  {title} {\enquote {\bibinfo
  {title} {Generation and manipulation of schrödinger cat states in rydberg
  atom arrays},}\ }\href {\doibase 10.1126/science.aax9743} {\bibfield
  {journal} {\bibinfo  {journal} {Science}\ }\textbf {\bibinfo {volume}
  {365}},\ \bibinfo {pages} {570--574} (\bibinfo {year} {2019})}\BibitemShut
  {NoStop}%
\bibitem [{\citenamefont {Levine}\ \emph {et~al.}(2019)\citenamefont {Levine},
  \citenamefont {Keesling}, \citenamefont {Semeghini}, \citenamefont {Omran},
  \citenamefont {Wang}, \citenamefont {Ebadi}, \citenamefont {Bernien},
  \citenamefont {Greiner}, \citenamefont {Vuleti\ifmmode~\acute{c}\else
  \'{c}\fi{}}, \citenamefont {Pichler},\ and\ \citenamefont
  {Lukin}}]{PhysRevLett.123.170503}%
  \BibitemOpen
  \bibfield  {author} {\bibinfo {author} {\bibfnamefont {Harry}\ \bibnamefont
  {Levine}}, \bibinfo {author} {\bibfnamefont {Alexander}\ \bibnamefont
  {Keesling}}, \bibinfo {author} {\bibfnamefont {Giulia}\ \bibnamefont
  {Semeghini}}, \bibinfo {author} {\bibfnamefont {Ahmed}\ \bibnamefont
  {Omran}}, \bibinfo {author} {\bibfnamefont {Tout~T.}\ \bibnamefont {Wang}},
  \bibinfo {author} {\bibfnamefont {Sepehr}\ \bibnamefont {Ebadi}}, \bibinfo
  {author} {\bibfnamefont {Hannes}\ \bibnamefont {Bernien}}, \bibinfo {author}
  {\bibfnamefont {Markus}\ \bibnamefont {Greiner}}, \bibinfo {author}
  {\bibfnamefont {Vladan}\ \bibnamefont {Vuleti\ifmmode~\acute{c}\else
  \'{c}\fi{}}}, \bibinfo {author} {\bibfnamefont {Hannes}\ \bibnamefont
  {Pichler}}, \ and\ \bibinfo {author} {\bibfnamefont {Mikhail~D.}\
  \bibnamefont {Lukin}},\ }\bibfield  {title} {\enquote {\bibinfo {title}
  {Parallel implementation of high-fidelity multiqubit gates with neutral
  atoms},}\ }\href {\doibase 10.1103/PhysRevLett.123.170503} {\bibfield
  {journal} {\bibinfo  {journal} {Phys. Rev. Lett.}\ }\textbf {\bibinfo
  {volume} {123}},\ \bibinfo {pages} {170503} (\bibinfo {year}
  {2019})}\BibitemShut {NoStop}%
\bibitem [{\citenamefont {Evered}\ \emph {et~al.}(2023)\citenamefont {Evered},
  \citenamefont {Bluvstein}, \citenamefont {Kalinowski}, \citenamefont {Ebadi},
  \citenamefont {Manovitz}, \citenamefont {Zhou}, \citenamefont {Li},
  \citenamefont {Geim}, \citenamefont {Wang}, \citenamefont {Maskara},
  \citenamefont {Levine}, \citenamefont {Semeghini}, \citenamefont {Greiner},
  \citenamefont {Vuleti{\'{c}}},\ and\ \citenamefont {Lukin}}]{Evered2023}%
  \BibitemOpen
  \bibfield  {author} {\bibinfo {author} {\bibfnamefont {Simon~J.}\
  \bibnamefont {Evered}}, \bibinfo {author} {\bibfnamefont {Dolev}\
  \bibnamefont {Bluvstein}}, \bibinfo {author} {\bibfnamefont {Marcin}\
  \bibnamefont {Kalinowski}}, \bibinfo {author} {\bibfnamefont {Sepehr}\
  \bibnamefont {Ebadi}}, \bibinfo {author} {\bibfnamefont {Tom}\ \bibnamefont
  {Manovitz}}, \bibinfo {author} {\bibfnamefont {Hengyun}\ \bibnamefont
  {Zhou}}, \bibinfo {author} {\bibfnamefont {Sophie~H.}\ \bibnamefont {Li}},
  \bibinfo {author} {\bibfnamefont {Alexandra~A.}\ \bibnamefont {Geim}},
  \bibinfo {author} {\bibfnamefont {Tout~T.}\ \bibnamefont {Wang}}, \bibinfo
  {author} {\bibfnamefont {Nishad}\ \bibnamefont {Maskara}}, \bibinfo {author}
  {\bibfnamefont {Harry}\ \bibnamefont {Levine}}, \bibinfo {author}
  {\bibfnamefont {Giulia}\ \bibnamefont {Semeghini}}, \bibinfo {author}
  {\bibfnamefont {Markus}\ \bibnamefont {Greiner}}, \bibinfo {author}
  {\bibfnamefont {Vladan}\ \bibnamefont {Vuleti{\'{c}}}}, \ and\ \bibinfo
  {author} {\bibfnamefont {Mikhail~D.}\ \bibnamefont {Lukin}},\ }\bibfield
  {title} {\enquote {\bibinfo {title} {High-fidelity parallel entangling gates
  on a neutral-atom quantum computer},}\ }\href {\doibase
  10.1038/s41586-023-06481-y} {\bibfield  {journal} {\bibinfo  {journal}
  {Nature}\ }\textbf {\bibinfo {volume} {622}},\ \bibinfo {pages} {268--272}
  (\bibinfo {year} {2023})}\BibitemShut {NoStop}%
\bibitem [{\citenamefont {Zhang}\ \emph {et~al.}(2024)\citenamefont {Zhang},
  \citenamefont {Zhang}, \citenamefont {Liu}, \citenamefont {Zhang},
  \citenamefont {Shao}, \citenamefont {Li}, \citenamefont {Chen}, \citenamefont
  {Liu}, \citenamefont {Ma}, \citenamefont {Han}, \citenamefont {Wang},
  \citenamefont {Adams}, \citenamefont {Shi},\ and\ \citenamefont
  {Ding}}]{PhysRevLett.133.243601}%
  \BibitemOpen
  \bibfield  {author} {\bibinfo {author} {\bibfnamefont {Jun}\ \bibnamefont
  {Zhang}}, \bibinfo {author} {\bibfnamefont {Li-Hua}\ \bibnamefont {Zhang}},
  \bibinfo {author} {\bibfnamefont {Bang}\ \bibnamefont {Liu}}, \bibinfo
  {author} {\bibfnamefont {Zheng-Yuan}\ \bibnamefont {Zhang}}, \bibinfo
  {author} {\bibfnamefont {Shi-Yao}\ \bibnamefont {Shao}}, \bibinfo {author}
  {\bibfnamefont {Qing}\ \bibnamefont {Li}}, \bibinfo {author} {\bibfnamefont
  {Han-Chao}\ \bibnamefont {Chen}}, \bibinfo {author} {\bibfnamefont
  {Zong-Kai}\ \bibnamefont {Liu}}, \bibinfo {author} {\bibfnamefont
  {Yu}~\bibnamefont {Ma}}, \bibinfo {author} {\bibfnamefont {Tian-Yu}\
  \bibnamefont {Han}}, \bibinfo {author} {\bibfnamefont {Qi-Feng}\ \bibnamefont
  {Wang}}, \bibinfo {author} {\bibfnamefont {C.~Stuart}\ \bibnamefont {Adams}},
  \bibinfo {author} {\bibfnamefont {Bao-Sen}\ \bibnamefont {Shi}}, \ and\
  \bibinfo {author} {\bibfnamefont {Dong-Sheng}\ \bibnamefont {Ding}},\
  }\bibfield  {title} {\enquote {\bibinfo {title} {Early warning signals of the
  tipping point in strongly interacting rydberg atoms},}\ }\href {\doibase
  10.1103/PhysRevLett.133.243601} {\bibfield  {journal} {\bibinfo  {journal}
  {Phys. Rev. Lett.}\ }\textbf {\bibinfo {volume} {133}},\ \bibinfo {pages}
  {243601} (\bibinfo {year} {2024})}\BibitemShut {NoStop}%
\bibitem [{\citenamefont {Anand}\ \emph {et~al.}(2024)\citenamefont {Anand},
  \citenamefont {Bradley}, \citenamefont {White}, \citenamefont {Ramesh},
  \citenamefont {Singh},\ and\ \citenamefont {Bernien}}]{Anand2024}%
  \BibitemOpen
  \bibfield  {author} {\bibinfo {author} {\bibfnamefont {Shraddha}\
  \bibnamefont {Anand}}, \bibinfo {author} {\bibfnamefont {Conor~E.}\
  \bibnamefont {Bradley}}, \bibinfo {author} {\bibfnamefont {Ryan}\
  \bibnamefont {White}}, \bibinfo {author} {\bibfnamefont {Vikram}\
  \bibnamefont {Ramesh}}, \bibinfo {author} {\bibfnamefont {Kevin}\
  \bibnamefont {Singh}}, \ and\ \bibinfo {author} {\bibfnamefont {Hannes}\
  \bibnamefont {Bernien}},\ }\bibfield  {title} {\enquote {\bibinfo {title} {A
  dual-species rydberg array},}\ }\href {\doibase 10.1038/s41567-024-02638-2}
  {\bibfield  {journal} {\bibinfo  {journal} {Nature Physics}\ }\textbf
  {\bibinfo {volume} {20}},\ \bibinfo {pages} {1744--1750} (\bibinfo {year}
  {2024})}\BibitemShut {NoStop}%
\bibitem [{\citenamefont {Crescimanna}\ \emph {et~al.}(2023)\citenamefont
  {Crescimanna}, \citenamefont {Taylor}, \citenamefont {Goldberg},\ and\
  \citenamefont {Heshami}}]{PhysRevApplied.20.034019}%
  \BibitemOpen
  \bibfield  {author} {\bibinfo {author} {\bibfnamefont {Valerio}\ \bibnamefont
  {Crescimanna}}, \bibinfo {author} {\bibfnamefont {Jacob}\ \bibnamefont
  {Taylor}}, \bibinfo {author} {\bibfnamefont {Aaron~Z.}\ \bibnamefont
  {Goldberg}}, \ and\ \bibinfo {author} {\bibfnamefont {Khabat}\ \bibnamefont
  {Heshami}},\ }\bibfield  {title} {\enquote {\bibinfo {title} {Quantum control
  of rydberg atoms for mesoscopic quantum state and circuit preparation},}\
  }\href {\doibase 10.1103/PhysRevApplied.20.034019} {\bibfield  {journal}
  {\bibinfo  {journal} {Phys. Rev. Appl.}\ }\textbf {\bibinfo {volume} {20}},\
  \bibinfo {pages} {034019} (\bibinfo {year} {2023})}\BibitemShut {NoStop}%
\bibitem [{\citenamefont {Jin}\ and\ \citenamefont
  {Jing}(2024)}]{PhysRevA.109.012619}%
  \BibitemOpen
  \bibfield  {author} {\bibinfo {author} {\bibfnamefont {Zhu-yao}\ \bibnamefont
  {Jin}}\ and\ \bibinfo {author} {\bibfnamefont {Jun}\ \bibnamefont {Jing}},\
  }\bibfield  {title} {\enquote {\bibinfo {title} {Geometric quantum gates via
  dark paths in rydberg atoms},}\ }\href {\doibase 10.1103/PhysRevA.109.012619}
  {\bibfield  {journal} {\bibinfo  {journal} {Phys. Rev. A}\ }\textbf {\bibinfo
  {volume} {109}},\ \bibinfo {pages} {012619} (\bibinfo {year}
  {2024})}\BibitemShut {NoStop}%
\bibitem [{\citenamefont {Wu}\ \emph {et~al.}(2023)\citenamefont {Wu},
  \citenamefont {Kirova}, \citenamefont {Auzins},\ and\ \citenamefont
  {Chen}}]{Wu:23}%
  \BibitemOpen
  \bibfield  {author} {\bibinfo {author} {\bibfnamefont {Chi-En}\ \bibnamefont
  {Wu}}, \bibinfo {author} {\bibfnamefont {Teodora}\ \bibnamefont {Kirova}},
  \bibinfo {author} {\bibfnamefont {Marcis}\ \bibnamefont {Auzins}}, \ and\
  \bibinfo {author} {\bibfnamefont {Yi-Hsin}\ \bibnamefont {Chen}},\ }\bibfield
   {title} {\enquote {\bibinfo {title} {Rydberg-rydberg interaction strengths
  and dipole blockade radii in the presence of f\"orster resonances},}\ }\href
  {\doibase 10.1364/OE.502183} {\bibfield  {journal} {\bibinfo  {journal} {Opt.
  Express}\ }\textbf {\bibinfo {volume} {31}},\ \bibinfo {pages} {37094--37104}
  (\bibinfo {year} {2023})}\BibitemShut {NoStop}%
\bibitem [{\citenamefont {Liu}\ \emph {et~al.}(2024)\citenamefont {Liu},
  \citenamefont {Zhang}, \citenamefont {Wang}, \citenamefont {Ma},
  \citenamefont {Han}, \citenamefont {Zhang}, \citenamefont {Zhang},
  \citenamefont {Shao}, \citenamefont {Li}, \citenamefont {Chen}, \citenamefont
  {Shi},\ and\ \citenamefont {Ding}}]{Liu2024}%
  \BibitemOpen
  \bibfield  {author} {\bibinfo {author} {\bibfnamefont {Bang}\ \bibnamefont
  {Liu}}, \bibinfo {author} {\bibfnamefont {Li-Hua}\ \bibnamefont {Zhang}},
  \bibinfo {author} {\bibfnamefont {Qi-Feng}\ \bibnamefont {Wang}}, \bibinfo
  {author} {\bibfnamefont {Yu}~\bibnamefont {Ma}}, \bibinfo {author}
  {\bibfnamefont {Tian-Yu}\ \bibnamefont {Han}}, \bibinfo {author}
  {\bibfnamefont {Jun}\ \bibnamefont {Zhang}}, \bibinfo {author} {\bibfnamefont
  {Zheng-Yuan}\ \bibnamefont {Zhang}}, \bibinfo {author} {\bibfnamefont
  {Shi-Yao}\ \bibnamefont {Shao}}, \bibinfo {author} {\bibfnamefont {Qing}\
  \bibnamefont {Li}}, \bibinfo {author} {\bibfnamefont {Han-Chao}\ \bibnamefont
  {Chen}}, \bibinfo {author} {\bibfnamefont {Bao-Sen}\ \bibnamefont {Shi}}, \
  and\ \bibinfo {author} {\bibfnamefont {Dong-Sheng}\ \bibnamefont {Ding}},\
  }\bibfield  {title} {\enquote {\bibinfo {title} {Higher-order and fractional
  discrete time crystals in floquet-driven rydberg atoms},}\ }\href {\doibase
  10.1038/s41467-024-53712-5} {\bibfield  {journal} {\bibinfo  {journal}
  {Nature Communications}\ }\textbf {\bibinfo {volume} {15}},\ \bibinfo {pages}
  {9730} (\bibinfo {year} {2024})}\BibitemShut {NoStop}%
\bibitem [{\citenamefont {Lukin}\ \emph {et~al.}(2001)\citenamefont {Lukin},
  \citenamefont {Fleischhauer}, \citenamefont {Cote}, \citenamefont {Duan},
  \citenamefont {Jaksch}, \citenamefont {Cirac},\ and\ \citenamefont
  {Zoller}}]{PhysRevLett.87.037901}%
  \BibitemOpen
  \bibfield  {author} {\bibinfo {author} {\bibfnamefont {M.~D.}\ \bibnamefont
  {Lukin}}, \bibinfo {author} {\bibfnamefont {M.}~\bibnamefont {Fleischhauer}},
  \bibinfo {author} {\bibfnamefont {R.}~\bibnamefont {Cote}}, \bibinfo {author}
  {\bibfnamefont {L.~M.}\ \bibnamefont {Duan}}, \bibinfo {author}
  {\bibfnamefont {D.}~\bibnamefont {Jaksch}}, \bibinfo {author} {\bibfnamefont
  {J.~I.}\ \bibnamefont {Cirac}}, \ and\ \bibinfo {author} {\bibfnamefont
  {P.}~\bibnamefont {Zoller}},\ }\bibfield  {title} {\enquote {\bibinfo {title}
  {Dipole blockade and quantum information processing in mesoscopic atomic
  ensembles},}\ }\href {\doibase 10.1103/PhysRevLett.87.037901} {\bibfield
  {journal} {\bibinfo  {journal} {Phys. Rev. Lett.}\ }\textbf {\bibinfo
  {volume} {87}},\ \bibinfo {pages} {037901} (\bibinfo {year}
  {2001})}\BibitemShut {NoStop}%
\bibitem [{\citenamefont {Urban}\ \emph {et~al.}(2009)\citenamefont {Urban},
  \citenamefont {Johnson}, \citenamefont {Henage}, \citenamefont {Isenhower},
  \citenamefont {Yavuz}, \citenamefont {Walker},\ and\ \citenamefont
  {Saffman}}]{Urban2009}%
  \BibitemOpen
  \bibfield  {author} {\bibinfo {author} {\bibfnamefont {E.}~\bibnamefont
  {Urban}}, \bibinfo {author} {\bibfnamefont {T.~A.}\ \bibnamefont {Johnson}},
  \bibinfo {author} {\bibfnamefont {T.}~\bibnamefont {Henage}}, \bibinfo
  {author} {\bibfnamefont {L.}~\bibnamefont {Isenhower}}, \bibinfo {author}
  {\bibfnamefont {D.~D.}\ \bibnamefont {Yavuz}}, \bibinfo {author}
  {\bibfnamefont {T.~G.}\ \bibnamefont {Walker}}, \ and\ \bibinfo {author}
  {\bibfnamefont {M.}~\bibnamefont {Saffman}},\ }\bibfield  {title} {\enquote
  {\bibinfo {title} {Observation of rydberg blockade between two atoms},}\
  }\href {\doibase 10.1038/nphys1178} {\bibfield  {journal} {\bibinfo
  {journal} {Nature Physics}\ }\textbf {\bibinfo {volume} {5}},\ \bibinfo
  {pages} {110--114} (\bibinfo {year} {2009})}\BibitemShut {NoStop}%
\bibitem [{\citenamefont {Graham}\ \emph {et~al.}(2019)\citenamefont {Graham},
  \citenamefont {Kwon}, \citenamefont {Grinkemeyer}, \citenamefont {Marra},
  \citenamefont {Jiang}, \citenamefont {Lichtman}, \citenamefont {Sun},
  \citenamefont {Ebert},\ and\ \citenamefont
  {Saffman}}]{PhysRevLett.123.230501}%
  \BibitemOpen
  \bibfield  {author} {\bibinfo {author} {\bibfnamefont {T.~M.}\ \bibnamefont
  {Graham}}, \bibinfo {author} {\bibfnamefont {M.}~\bibnamefont {Kwon}},
  \bibinfo {author} {\bibfnamefont {B.}~\bibnamefont {Grinkemeyer}}, \bibinfo
  {author} {\bibfnamefont {Z.}~\bibnamefont {Marra}}, \bibinfo {author}
  {\bibfnamefont {X.}~\bibnamefont {Jiang}}, \bibinfo {author} {\bibfnamefont
  {M.~T.}\ \bibnamefont {Lichtman}}, \bibinfo {author} {\bibfnamefont
  {Y.}~\bibnamefont {Sun}}, \bibinfo {author} {\bibfnamefont {M.}~\bibnamefont
  {Ebert}}, \ and\ \bibinfo {author} {\bibfnamefont {M.}~\bibnamefont
  {Saffman}},\ }\bibfield  {title} {\enquote {\bibinfo {title}
  {Rydberg-mediated entanglement in a two-dimensional neutral atom qubit
  array},}\ }\href {\doibase 10.1103/PhysRevLett.123.230501} {\bibfield
  {journal} {\bibinfo  {journal} {Phys. Rev. Lett.}\ }\textbf {\bibinfo
  {volume} {123}},\ \bibinfo {pages} {230501} (\bibinfo {year}
  {2019})}\BibitemShut {NoStop}%
\bibitem [{\citenamefont {Fromonteil}\ \emph {et~al.}(2023)\citenamefont
  {Fromonteil}, \citenamefont {Bluvstein},\ and\ \citenamefont
  {Pichler}}]{PRXQuantum.4.020335}%
  \BibitemOpen
  \bibfield  {author} {\bibinfo {author} {\bibfnamefont {Charles}\ \bibnamefont
  {Fromonteil}}, \bibinfo {author} {\bibfnamefont {Dolev}\ \bibnamefont
  {Bluvstein}}, \ and\ \bibinfo {author} {\bibfnamefont {Hannes}\ \bibnamefont
  {Pichler}},\ }\bibfield  {title} {\enquote {\bibinfo {title} {Protocols for
  rydberg entangling gates featuring robustness against quasistatic errors},}\
  }\href {\doibase 10.1103/PRXQuantum.4.020335} {\bibfield  {journal} {\bibinfo
   {journal} {PRX Quantum}\ }\textbf {\bibinfo {volume} {4}},\ \bibinfo {pages}
  {020335} (\bibinfo {year} {2023})}\BibitemShut {NoStop}%
\bibitem [{\citenamefont {Buchemmavari}\ \emph {et~al.}(2024)\citenamefont
  {Buchemmavari}, \citenamefont {Omanakuttan}, \citenamefont {Jau},\ and\
  \citenamefont {Deutsch}}]{PhysRevA.109.012615}%
  \BibitemOpen
  \bibfield  {author} {\bibinfo {author} {\bibfnamefont {Vikas}\ \bibnamefont
  {Buchemmavari}}, \bibinfo {author} {\bibfnamefont {Sivaprasad}\ \bibnamefont
  {Omanakuttan}}, \bibinfo {author} {\bibfnamefont {Yuan-Yu}\ \bibnamefont
  {Jau}}, \ and\ \bibinfo {author} {\bibfnamefont {Ivan}\ \bibnamefont
  {Deutsch}},\ }\bibfield  {title} {\enquote {\bibinfo {title} {Entangling
  quantum logic gates in neutral atoms via the microwave-driven spin-flip
  blockade},}\ }\href {\doibase 10.1103/PhysRevA.109.012615} {\bibfield
  {journal} {\bibinfo  {journal} {Phys. Rev. A}\ }\textbf {\bibinfo {volume}
  {109}},\ \bibinfo {pages} {012615} (\bibinfo {year} {2024})}\BibitemShut
  {NoStop}%
\bibitem [{\citenamefont {Sun}(2024)}]{Sun2024}%
  \BibitemOpen
  \bibfield  {author} {\bibinfo {author} {\bibfnamefont {Yuan}\ \bibnamefont
  {Sun}},\ }\bibfield  {title} {\enquote {\bibinfo {title}
  {Buffer-atom-mediated quantum logic gates with off-resonant modulated
  driving},}\ }\href {\doibase 10.1007/s11433-024-2478-8} {\bibfield  {journal}
  {\bibinfo  {journal} {Science China Physics, Mechanics {\&} Astronomy}\
  }\textbf {\bibinfo {volume} {67}},\ \bibinfo {pages} {120311} (\bibinfo
  {year} {2024})}\BibitemShut {NoStop}%
\bibitem [{\citenamefont {Chen}\ \emph {et~al.}(2024)\citenamefont {Chen},
  \citenamefont {Liang}, \citenamefont {Fu}, \citenamefont {Liu}, \citenamefont
  {He}, \citenamefont {Wang}, \citenamefont {Han}, \citenamefont {Huang},
  \citenamefont {Lv},\ and\ \citenamefont {Du}}]{PhysRevA.109.042621}%
  \BibitemOpen
  \bibfield  {author} {\bibinfo {author} {\bibfnamefont {Zi-Yuan}\ \bibnamefont
  {Chen}}, \bibinfo {author} {\bibfnamefont {Jia-Hao}\ \bibnamefont {Liang}},
  \bibinfo {author} {\bibfnamefont {Zhao-Xin}\ \bibnamefont {Fu}}, \bibinfo
  {author} {\bibfnamefont {Hong-Zhi}\ \bibnamefont {Liu}}, \bibinfo {author}
  {\bibfnamefont {Ze-Rui}\ \bibnamefont {He}}, \bibinfo {author} {\bibfnamefont
  {Meng}\ \bibnamefont {Wang}}, \bibinfo {author} {\bibfnamefont {Zhi-Wei}\
  \bibnamefont {Han}}, \bibinfo {author} {\bibfnamefont {Jia-Yi}\ \bibnamefont
  {Huang}}, \bibinfo {author} {\bibfnamefont {Qing-Xian}\ \bibnamefont {Lv}}, \
  and\ \bibinfo {author} {\bibfnamefont {Yan-Xiong}\ \bibnamefont {Du}},\
  }\bibfield  {title} {\enquote {\bibinfo {title} {Single-pulse two-qubit gates
  for rydberg atoms with noncyclic geometric control},}\ }\href {\doibase
  10.1103/PhysRevA.109.042621} {\bibfield  {journal} {\bibinfo  {journal}
  {Phys. Rev. A}\ }\textbf {\bibinfo {volume} {109}},\ \bibinfo {pages}
  {042621} (\bibinfo {year} {2024})}\BibitemShut {NoStop}%
\bibitem [{\citenamefont {Isenhower}\ \emph {et~al.}(2010)\citenamefont
  {Isenhower}, \citenamefont {Urban}, \citenamefont {Zhang}, \citenamefont
  {Gill}, \citenamefont {Henage}, \citenamefont {Johnson}, \citenamefont
  {Walker},\ and\ \citenamefont {Saffman}}]{PhysRevLett.104.010503}%
  \BibitemOpen
  \bibfield  {author} {\bibinfo {author} {\bibfnamefont {L.}~\bibnamefont
  {Isenhower}}, \bibinfo {author} {\bibfnamefont {E.}~\bibnamefont {Urban}},
  \bibinfo {author} {\bibfnamefont {X.~L.}\ \bibnamefont {Zhang}}, \bibinfo
  {author} {\bibfnamefont {A.~T.}\ \bibnamefont {Gill}}, \bibinfo {author}
  {\bibfnamefont {T.}~\bibnamefont {Henage}}, \bibinfo {author} {\bibfnamefont
  {T.~A.}\ \bibnamefont {Johnson}}, \bibinfo {author} {\bibfnamefont {T.~G.}\
  \bibnamefont {Walker}}, \ and\ \bibinfo {author} {\bibfnamefont
  {M.}~\bibnamefont {Saffman}},\ }\bibfield  {title} {\enquote {\bibinfo
  {title} {Demonstration of a neutral atom controlled-not quantum gate},}\
  }\href {\doibase 10.1103/PhysRevLett.104.010503} {\bibfield  {journal}
  {\bibinfo  {journal} {Phys. Rev. Lett.}\ }\textbf {\bibinfo {volume} {104}},\
  \bibinfo {pages} {010503} (\bibinfo {year} {2010})}\BibitemShut {NoStop}%
\bibitem [{\citenamefont {Su}\ \emph {et~al.}(2023)\citenamefont {Su},
  \citenamefont {Sun}, \citenamefont {Liu}, \citenamefont {Yan}, \citenamefont
  {Yung}, \citenamefont {Li},\ and\ \citenamefont
  {Feng}}]{PhysRevApplied.19.044007}%
  \BibitemOpen
  \bibfield  {author} {\bibinfo {author} {\bibfnamefont {S.-L.}\ \bibnamefont
  {Su}}, \bibinfo {author} {\bibfnamefont {Li-Na}\ \bibnamefont {Sun}},
  \bibinfo {author} {\bibfnamefont {B.-J.}\ \bibnamefont {Liu}}, \bibinfo
  {author} {\bibfnamefont {L.-L.}\ \bibnamefont {Yan}}, \bibinfo {author}
  {\bibfnamefont {M.-H.}\ \bibnamefont {Yung}}, \bibinfo {author}
  {\bibfnamefont {W.}~\bibnamefont {Li}}, \ and\ \bibinfo {author}
  {\bibfnamefont {M.}~\bibnamefont {Feng}},\ }\bibfield  {title} {\enquote
  {\bibinfo {title} {Rabi- and blockade-error-resilient all-geometric rydberg
  quantum gates},}\ }\href {\doibase 10.1103/PhysRevApplied.19.044007}
  {\bibfield  {journal} {\bibinfo  {journal} {Phys. Rev. Appl.}\ }\textbf
  {\bibinfo {volume} {19}},\ \bibinfo {pages} {044007} (\bibinfo {year}
  {2023})}\BibitemShut {NoStop}%
\bibitem [{\citenamefont {M\o{}ller}\ \emph {et~al.}(2008)\citenamefont
  {M\o{}ller}, \citenamefont {Madsen},\ and\ \citenamefont
  {M\o{}lmer}}]{PhysRevLett.100.170504}%
  \BibitemOpen
  \bibfield  {author} {\bibinfo {author} {\bibfnamefont {Ditte}\ \bibnamefont
  {M\o{}ller}}, \bibinfo {author} {\bibfnamefont {Lars~Bojer}\ \bibnamefont
  {Madsen}}, \ and\ \bibinfo {author} {\bibfnamefont {Klaus}\ \bibnamefont
  {M\o{}lmer}},\ }\bibfield  {title} {\enquote {\bibinfo {title} {Quantum gates
  and multiparticle entanglement by rydberg excitation blockade and adiabatic
  passage},}\ }\href {\doibase 10.1103/PhysRevLett.100.170504} {\bibfield
  {journal} {\bibinfo  {journal} {Phys. Rev. Lett.}\ }\textbf {\bibinfo
  {volume} {100}},\ \bibinfo {pages} {170504} (\bibinfo {year}
  {2008})}\BibitemShut {NoStop}%
\bibitem [{\citenamefont {Liu}\ \emph {et~al.}(2020)\citenamefont {Liu},
  \citenamefont {Su},\ and\ \citenamefont {Yung}}]{PhysRevResearch.2.043130}%
  \BibitemOpen
  \bibfield  {author} {\bibinfo {author} {\bibfnamefont {Bao-Jie}\ \bibnamefont
  {Liu}}, \bibinfo {author} {\bibfnamefont {Shi-Lei}\ \bibnamefont {Su}}, \
  and\ \bibinfo {author} {\bibfnamefont {Man-Hong}\ \bibnamefont {Yung}},\
  }\bibfield  {title} {\enquote {\bibinfo {title} {Nonadiabatic noncyclic
  geometric quantum computation in rydberg atoms},}\ }\href {\doibase
  10.1103/PhysRevResearch.2.043130} {\bibfield  {journal} {\bibinfo  {journal}
  {Phys. Rev. Res.}\ }\textbf {\bibinfo {volume} {2}},\ \bibinfo {pages}
  {043130} (\bibinfo {year} {2020})}\BibitemShut {NoStop}%
\bibitem [{\citenamefont {Xue}\ \emph {et~al.}(2024)\citenamefont {Xue},
  \citenamefont {Xu}, \citenamefont {Li},\ and\ \citenamefont
  {Li}}]{PhysRevA.110.032619}%
  \BibitemOpen
  \bibfield  {author} {\bibinfo {author} {\bibfnamefont {Ming}\ \bibnamefont
  {Xue}}, \bibinfo {author} {\bibfnamefont {Shijie}\ \bibnamefont {Xu}},
  \bibinfo {author} {\bibfnamefont {Xinwei}\ \bibnamefont {Li}}, \ and\
  \bibinfo {author} {\bibfnamefont {Xiangliang}\ \bibnamefont {Li}},\
  }\bibfield  {title} {\enquote {\bibinfo {title} {High-fidelity and robust
  controlled-$z$ gates implemented with rydberg atoms via echoing rapid
  adiabatic passage},}\ }\href {\doibase 10.1103/PhysRevA.110.032619}
  {\bibfield  {journal} {\bibinfo  {journal} {Phys. Rev. A}\ }\textbf {\bibinfo
  {volume} {110}},\ \bibinfo {pages} {032619} (\bibinfo {year}
  {2024})}\BibitemShut {NoStop}%
\bibitem [{\citenamefont {Guo}\ \emph {et~al.}(2020)\citenamefont {Guo},
  \citenamefont {Wu}, \citenamefont {Zhu}, \citenamefont {Jin}, \citenamefont
  {Zeng}, \citenamefont {Zhang}, \citenamefont {Yan}, \citenamefont {Feng},\
  and\ \citenamefont {Su}}]{PhysRevA.102.062410}%
  \BibitemOpen
  \bibfield  {author} {\bibinfo {author} {\bibfnamefont {F.-Q.}\ \bibnamefont
  {Guo}}, \bibinfo {author} {\bibfnamefont {J.-L.}\ \bibnamefont {Wu}},
  \bibinfo {author} {\bibfnamefont {X.-Y.}\ \bibnamefont {Zhu}}, \bibinfo
  {author} {\bibfnamefont {Z.}~\bibnamefont {Jin}}, \bibinfo {author}
  {\bibfnamefont {Y.}~\bibnamefont {Zeng}}, \bibinfo {author} {\bibfnamefont
  {S.}~\bibnamefont {Zhang}}, \bibinfo {author} {\bibfnamefont {L.-L.}\
  \bibnamefont {Yan}}, \bibinfo {author} {\bibfnamefont {M.}~\bibnamefont
  {Feng}}, \ and\ \bibinfo {author} {\bibfnamefont {S.-L.}\ \bibnamefont
  {Su}},\ }\bibfield  {title} {\enquote {\bibinfo {title} {Complete and
  nondestructive distinguishment of many-body rydberg entanglement via robust
  geometric quantum operations},}\ }\href {\doibase
  10.1103/PhysRevA.102.062410} {\bibfield  {journal} {\bibinfo  {journal}
  {Phys. Rev. A}\ }\textbf {\bibinfo {volume} {102}},\ \bibinfo {pages}
  {062410} (\bibinfo {year} {2020})}\BibitemShut {NoStop}%
\bibitem [{\citenamefont {Mitra}\ \emph {et~al.}(2023)\citenamefont {Mitra},
  \citenamefont {Omanakuttan}, \citenamefont {Martin}, \citenamefont
  {Biedermann},\ and\ \citenamefont {Deutsch}}]{PhysRevA.107.062609}%
  \BibitemOpen
  \bibfield  {author} {\bibinfo {author} {\bibfnamefont {Anupam}\ \bibnamefont
  {Mitra}}, \bibinfo {author} {\bibfnamefont {Sivaprasad}\ \bibnamefont
  {Omanakuttan}}, \bibinfo {author} {\bibfnamefont {Michael~J.}\ \bibnamefont
  {Martin}}, \bibinfo {author} {\bibfnamefont {Grant~W.}\ \bibnamefont
  {Biedermann}}, \ and\ \bibinfo {author} {\bibfnamefont {Ivan~H.}\
  \bibnamefont {Deutsch}},\ }\bibfield  {title} {\enquote {\bibinfo {title}
  {Neutral-atom entanglement using adiabatic rydberg dressing},}\ }\href
  {\doibase 10.1103/PhysRevA.107.062609} {\bibfield  {journal} {\bibinfo
  {journal} {Phys. Rev. A}\ }\textbf {\bibinfo {volume} {107}},\ \bibinfo
  {pages} {062609} (\bibinfo {year} {2023})}\BibitemShut {NoStop}%
\bibitem [{\citenamefont {Jandura}\ \emph {et~al.}(2023)\citenamefont
  {Jandura}, \citenamefont {Thompson},\ and\ \citenamefont
  {Pupillo}}]{PRXQuantum.4.020336}%
  \BibitemOpen
  \bibfield  {author} {\bibinfo {author} {\bibfnamefont {Sven}\ \bibnamefont
  {Jandura}}, \bibinfo {author} {\bibfnamefont {Jeff~D.}\ \bibnamefont
  {Thompson}}, \ and\ \bibinfo {author} {\bibfnamefont {Guido}\ \bibnamefont
  {Pupillo}},\ }\bibfield  {title} {\enquote {\bibinfo {title} {Optimizing
  rydberg gates for logical-qubit performance},}\ }\href {\doibase
  10.1103/PRXQuantum.4.020336} {\bibfield  {journal} {\bibinfo  {journal} {PRX
  Quantum}\ }\textbf {\bibinfo {volume} {4}},\ \bibinfo {pages} {020336}
  (\bibinfo {year} {2023})}\BibitemShut {NoStop}%
\bibitem [{\citenamefont {Song}\ \emph {et~al.}(2024)\citenamefont {Song},
  \citenamefont {Wei}, \citenamefont {Xu}, \citenamefont {Yan}, \citenamefont
  {Feng}, \citenamefont {Su},\ and\ \citenamefont
  {Chen}}]{PhysRevA.109.022613}%
  \BibitemOpen
  \bibfield  {author} {\bibinfo {author} {\bibfnamefont {P.-Y.}\ \bibnamefont
  {Song}}, \bibinfo {author} {\bibfnamefont {J.-F.}\ \bibnamefont {Wei}},
  \bibinfo {author} {\bibfnamefont {Peng}\ \bibnamefont {Xu}}, \bibinfo
  {author} {\bibfnamefont {L.-L.}\ \bibnamefont {Yan}}, \bibinfo {author}
  {\bibfnamefont {M.}~\bibnamefont {Feng}}, \bibinfo {author} {\bibfnamefont
  {Shi-Lei}\ \bibnamefont {Su}}, \ and\ \bibinfo {author} {\bibfnamefont
  {Gang}\ \bibnamefont {Chen}},\ }\bibfield  {title} {\enquote {\bibinfo
  {title} {Fast realization of high-fidelity nonadiabatic holonomic quantum
  gates with a time-optimal-control technique in rydberg atoms},}\ }\href
  {\doibase 10.1103/PhysRevA.109.022613} {\bibfield  {journal} {\bibinfo
  {journal} {Phys. Rev. A}\ }\textbf {\bibinfo {volume} {109}},\ \bibinfo
  {pages} {022613} (\bibinfo {year} {2024})}\BibitemShut {NoStop}%
\bibitem [{\citenamefont {Cao}\ \emph {et~al.}(2024)\citenamefont {Cao},
  \citenamefont {Eckner}, \citenamefont {Lukin~Yelin}, \citenamefont {Young},
  \citenamefont {Jandura}, \citenamefont {Yan}, \citenamefont {Kim},
  \citenamefont {Pupillo}, \citenamefont {Ye}, \citenamefont {Darkwah~Oppong},\
  and\ \citenamefont {Kaufman}}]{Cao2024}%
  \BibitemOpen
  \bibfield  {author} {\bibinfo {author} {\bibfnamefont {Alec}\ \bibnamefont
  {Cao}}, \bibinfo {author} {\bibfnamefont {William~J.}\ \bibnamefont
  {Eckner}}, \bibinfo {author} {\bibfnamefont {Theodor}\ \bibnamefont
  {Lukin~Yelin}}, \bibinfo {author} {\bibfnamefont {Aaron~W.}\ \bibnamefont
  {Young}}, \bibinfo {author} {\bibfnamefont {Sven}\ \bibnamefont {Jandura}},
  \bibinfo {author} {\bibfnamefont {Lingfeng}\ \bibnamefont {Yan}}, \bibinfo
  {author} {\bibfnamefont {Kyungtae}\ \bibnamefont {Kim}}, \bibinfo {author}
  {\bibfnamefont {Guido}\ \bibnamefont {Pupillo}}, \bibinfo {author}
  {\bibfnamefont {Jun}\ \bibnamefont {Ye}}, \bibinfo {author} {\bibfnamefont
  {Nelson}\ \bibnamefont {Darkwah~Oppong}}, \ and\ \bibinfo {author}
  {\bibfnamefont {Adam~M.}\ \bibnamefont {Kaufman}},\ }\bibfield  {title}
  {\enquote {\bibinfo {title} {Multi-qubit gates and schr{\"o}dinger cat states
  in an optical clock},}\ }\href {\doibase 10.1038/s41586-024-07913-z}
  {\bibfield  {journal} {\bibinfo  {journal} {Nature}\ }\textbf {\bibinfo
  {volume} {634}},\ \bibinfo {pages} {315--320} (\bibinfo {year}
  {2024})}\BibitemShut {NoStop}%
\bibitem [{\citenamefont {Jia}\ \emph {et~al.}(2024)\citenamefont {Jia},
  \citenamefont {Huie}, \citenamefont {Li}, \citenamefont {Sun}, \citenamefont
  {Hu}, \citenamefont {{Aakash}}, \citenamefont {Kogan}, \citenamefont {Karve},
  \citenamefont {Lee},\ and\ \citenamefont {Covey}}]{Jia2024}%
  \BibitemOpen
  \bibfield  {author} {\bibinfo {author} {\bibfnamefont {Zhubing}\ \bibnamefont
  {Jia}}, \bibinfo {author} {\bibfnamefont {William}\ \bibnamefont {Huie}},
  \bibinfo {author} {\bibfnamefont {Lintao}\ \bibnamefont {Li}}, \bibinfo
  {author} {\bibfnamefont {Won Kyu~Calvin}\ \bibnamefont {Sun}}, \bibinfo
  {author} {\bibfnamefont {Xiye}\ \bibnamefont {Hu}}, \bibinfo {author}
  {\bibnamefont {{Aakash}}}, \bibinfo {author} {\bibfnamefont {Healey}\
  \bibnamefont {Kogan}}, \bibinfo {author} {\bibfnamefont {Abhishek}\
  \bibnamefont {Karve}}, \bibinfo {author} {\bibfnamefont {Jong~Yeon}\
  \bibnamefont {Lee}}, \ and\ \bibinfo {author} {\bibfnamefont {Jacob~P.}\
  \bibnamefont {Covey}},\ }\bibfield  {title} {\enquote {\bibinfo {title} {An
  architecture for two-qubit encoding in neutral ytterbium-171 atoms},}\ }\href
  {\doibase 10.1038/s41534-024-00898-7} {\bibfield  {journal} {\bibinfo
  {journal} {npj Quantum Information}\ }\textbf {\bibinfo {volume} {10}},\
  \bibinfo {pages} {106} (\bibinfo {year} {2024})}\BibitemShut {NoStop}%
\bibitem [{\citenamefont {Unnikrishnan}\ \emph {et~al.}(2024)\citenamefont
  {Unnikrishnan}, \citenamefont {Ilzh\"ofer}, \citenamefont {Scholz},
  \citenamefont {H\"olzl}, \citenamefont {G\"otzelmann}, \citenamefont {Gupta},
  \citenamefont {Zhao}, \citenamefont {Krauter}, \citenamefont {Weber},
  \citenamefont {Makki}, \citenamefont {B\"uchler}, \citenamefont {Pfau},\ and\
  \citenamefont {Meinert}}]{PhysRevLett.132.150606}%
  \BibitemOpen
  \bibfield  {author} {\bibinfo {author} {\bibfnamefont {G.}~\bibnamefont
  {Unnikrishnan}}, \bibinfo {author} {\bibfnamefont {P.}~\bibnamefont
  {Ilzh\"ofer}}, \bibinfo {author} {\bibfnamefont {A.}~\bibnamefont {Scholz}},
  \bibinfo {author} {\bibfnamefont {C.}~\bibnamefont {H\"olzl}}, \bibinfo
  {author} {\bibfnamefont {A.}~\bibnamefont {G\"otzelmann}}, \bibinfo {author}
  {\bibfnamefont {R.~K.}\ \bibnamefont {Gupta}}, \bibinfo {author}
  {\bibfnamefont {J.}~\bibnamefont {Zhao}}, \bibinfo {author} {\bibfnamefont
  {J.}~\bibnamefont {Krauter}}, \bibinfo {author} {\bibfnamefont
  {S.}~\bibnamefont {Weber}}, \bibinfo {author} {\bibfnamefont
  {N.}~\bibnamefont {Makki}}, \bibinfo {author} {\bibfnamefont {H.~P.}\
  \bibnamefont {B\"uchler}}, \bibinfo {author} {\bibfnamefont {T.}~\bibnamefont
  {Pfau}}, \ and\ \bibinfo {author} {\bibfnamefont {F.}~\bibnamefont
  {Meinert}},\ }\bibfield  {title} {\enquote {\bibinfo {title} {Coherent
  control of the fine-structure qubit in a single alkaline-earth atom},}\
  }\href {\doibase 10.1103/PhysRevLett.132.150606} {\bibfield  {journal}
  {\bibinfo  {journal} {Phys. Rev. Lett.}\ }\textbf {\bibinfo {volume} {132}},\
  \bibinfo {pages} {150606} (\bibinfo {year} {2024})}\BibitemShut {NoStop}%
\bibitem [{\citenamefont {Stiesdal}\ \emph {et~al.}(2018)\citenamefont
  {Stiesdal}, \citenamefont {Kumlin}, \citenamefont {Kleinbeck}, \citenamefont
  {Lunt}, \citenamefont {Braun}, \citenamefont {Paris-Mandoki}, \citenamefont
  {Tresp}, \citenamefont {B\"uchler},\ and\ \citenamefont
  {Hofferberth}}]{PhysRevLett.121.103601}%
  \BibitemOpen
  \bibfield  {author} {\bibinfo {author} {\bibfnamefont {Nina}\ \bibnamefont
  {Stiesdal}}, \bibinfo {author} {\bibfnamefont {Jan}\ \bibnamefont {Kumlin}},
  \bibinfo {author} {\bibfnamefont {Kevin}\ \bibnamefont {Kleinbeck}}, \bibinfo
  {author} {\bibfnamefont {Philipp}\ \bibnamefont {Lunt}}, \bibinfo {author}
  {\bibfnamefont {Christoph}\ \bibnamefont {Braun}}, \bibinfo {author}
  {\bibfnamefont {Asaf}\ \bibnamefont {Paris-Mandoki}}, \bibinfo {author}
  {\bibfnamefont {Christoph}\ \bibnamefont {Tresp}}, \bibinfo {author}
  {\bibfnamefont {Hans~Peter}\ \bibnamefont {B\"uchler}}, \ and\ \bibinfo
  {author} {\bibfnamefont {Sebastian}\ \bibnamefont {Hofferberth}},\ }\bibfield
   {title} {\enquote {\bibinfo {title} {Observation of three-body correlations
  for photons coupled to a rydberg superatom},}\ }\href {\doibase
  10.1103/PhysRevLett.121.103601} {\bibfield  {journal} {\bibinfo  {journal}
  {Phys. Rev. Lett.}\ }\textbf {\bibinfo {volume} {121}},\ \bibinfo {pages}
  {103601} (\bibinfo {year} {2018})}\BibitemShut {NoStop}%
\bibitem [{\citenamefont {Zeiher}\ \emph {et~al.}(2015)\citenamefont {Zeiher},
  \citenamefont {Schau\ss{}}, \citenamefont {Hild}, \citenamefont {Macr\`{\i}},
  \citenamefont {Bloch},\ and\ \citenamefont {Gross}}]{PhysRevX.5.031015}%
  \BibitemOpen
  \bibfield  {author} {\bibinfo {author} {\bibfnamefont {Johannes}\
  \bibnamefont {Zeiher}}, \bibinfo {author} {\bibfnamefont {Peter}\
  \bibnamefont {Schau\ss{}}}, \bibinfo {author} {\bibfnamefont {Sebastian}\
  \bibnamefont {Hild}}, \bibinfo {author} {\bibfnamefont {Tommaso}\
  \bibnamefont {Macr\`{\i}}}, \bibinfo {author} {\bibfnamefont {Immanuel}\
  \bibnamefont {Bloch}}, \ and\ \bibinfo {author} {\bibfnamefont {Christian}\
  \bibnamefont {Gross}},\ }\bibfield  {title} {\enquote {\bibinfo {title}
  {Microscopic characterization of scalable coherent rydberg superatoms},}\
  }\href {\doibase 10.1103/PhysRevX.5.031015} {\bibfield  {journal} {\bibinfo
  {journal} {Phys. Rev. X}\ }\textbf {\bibinfo {volume} {5}},\ \bibinfo {pages}
  {031015} (\bibinfo {year} {2015})}\BibitemShut {NoStop}%
\bibitem [{\citenamefont {Shao}\ \emph {et~al.}(2024)\citenamefont {Shao},
  \citenamefont {Su}, \citenamefont {Li}, \citenamefont {Nath}, \citenamefont
  {Wu},\ and\ \citenamefont {Li}}]{Shao2024}%
  \BibitemOpen
  \bibfield  {author} {\bibinfo {author} {\bibfnamefont {Xiao-Qiang}\
  \bibnamefont {Shao}}, \bibinfo {author} {\bibfnamefont {Shi-Lei}\
  \bibnamefont {Su}}, \bibinfo {author} {\bibfnamefont {Lin}\ \bibnamefont
  {Li}}, \bibinfo {author} {\bibfnamefont {Rejish}\ \bibnamefont {Nath}},
  \bibinfo {author} {\bibfnamefont {Jin-Hui}\ \bibnamefont {Wu}}, \ and\
  \bibinfo {author} {\bibfnamefont {Weibin}\ \bibnamefont {Li}},\ }\bibfield
  {title} {\enquote {\bibinfo {title} {Rydberg superatoms: An artificial
  quantum system for quantum information processing and quantum optics},}\
  }\href {\doibase 10.1063/5.0211071} {\bibfield  {journal} {\bibinfo
  {journal} {Applied Physics Reviews}\ }\textbf {\bibinfo {volume} {11}},\
  \bibinfo {pages} {031320} (\bibinfo {year} {2024})}\BibitemShut {NoStop}%
\bibitem [{\citenamefont {Ates}\ \emph {et~al.}(2007)\citenamefont {Ates},
  \citenamefont {Pohl}, \citenamefont {Pattard},\ and\ \citenamefont
  {Rost}}]{Ates2007Antiblockade}%
  \BibitemOpen
  \bibfield  {author} {\bibinfo {author} {\bibfnamefont {C.}~\bibnamefont
  {Ates}}, \bibinfo {author} {\bibfnamefont {T.}~\bibnamefont {Pohl}}, \bibinfo
  {author} {\bibfnamefont {T.}~\bibnamefont {Pattard}}, \ and\ \bibinfo
  {author} {\bibfnamefont {J.~M.}\ \bibnamefont {Rost}},\ }\bibfield  {title}
  {\enquote {\bibinfo {title} {Antiblockade in rydberg excitation of an
  ultracold lattice gas},}\ }\href {\doibase 10.1103/PhysRevLett.98.023002}
  {\bibfield  {journal} {\bibinfo  {journal} {Phys. Rev. Lett.}\ }\textbf
  {\bibinfo {volume} {98}},\ \bibinfo {pages} {023002} (\bibinfo {year}
  {2007})}\BibitemShut {NoStop}%
\bibitem [{\citenamefont {Rao}\ and\ \citenamefont
  {M\o{}lmer}(2013)}]{PhysRevLett.111.033606}%
  \BibitemOpen
  \bibfield  {author} {\bibinfo {author} {\bibfnamefont {D.~D.~Bhaktavatsala}\
  \bibnamefont {Rao}}\ and\ \bibinfo {author} {\bibfnamefont {Klaus}\
  \bibnamefont {M\o{}lmer}},\ }\bibfield  {title} {\enquote {\bibinfo {title}
  {Dark entangled steady states of interacting rydberg atoms},}\ }\href
  {\doibase 10.1103/PhysRevLett.111.033606} {\bibfield  {journal} {\bibinfo
  {journal} {Phys. Rev. Lett.}\ }\textbf {\bibinfo {volume} {111}},\ \bibinfo
  {pages} {033606} (\bibinfo {year} {2013})}\BibitemShut {NoStop}%
\bibitem [{\citenamefont {Carr}\ and\ \citenamefont
  {Saffman}(2013)}]{PhysRevLett.111.033607}%
  \BibitemOpen
  \bibfield  {author} {\bibinfo {author} {\bibfnamefont {A.~W.}\ \bibnamefont
  {Carr}}\ and\ \bibinfo {author} {\bibfnamefont {M.}~\bibnamefont {Saffman}},\
  }\bibfield  {title} {\enquote {\bibinfo {title} {Preparation of entangled and
  antiferromagnetic states by dissipative rydberg pumping},}\ }\href {\doibase
  10.1103/PhysRevLett.111.033607} {\bibfield  {journal} {\bibinfo  {journal}
  {Phys. Rev. Lett.}\ }\textbf {\bibinfo {volume} {111}},\ \bibinfo {pages}
  {033607} (\bibinfo {year} {2013})}\BibitemShut {NoStop}%
\bibitem [{\citenamefont {Amthor}\ \emph {et~al.}(2010)\citenamefont {Amthor},
  \citenamefont {Giese}, \citenamefont {Hofmann},\ and\ \citenamefont
  {Weidem\"uller}}]{PhysRevLett.104.013001}%
  \BibitemOpen
  \bibfield  {author} {\bibinfo {author} {\bibfnamefont {Thomas}\ \bibnamefont
  {Amthor}}, \bibinfo {author} {\bibfnamefont {Christian}\ \bibnamefont
  {Giese}}, \bibinfo {author} {\bibfnamefont {Christoph~S.}\ \bibnamefont
  {Hofmann}}, \ and\ \bibinfo {author} {\bibfnamefont {Matthias}\ \bibnamefont
  {Weidem\"uller}},\ }\bibfield  {title} {\enquote {\bibinfo {title} {Evidence
  of antiblockade in an ultracold rydberg gas},}\ }\href {\doibase
  10.1103/PhysRevLett.104.013001} {\bibfield  {journal} {\bibinfo  {journal}
  {Phys. Rev. Lett.}\ }\textbf {\bibinfo {volume} {104}},\ \bibinfo {pages}
  {013001} (\bibinfo {year} {2010})}\BibitemShut {NoStop}%
\bibitem [{\citenamefont {Liu}\ \emph {et~al.}(2022{\natexlab{a}})\citenamefont
  {Liu}, \citenamefont {Yang}, \citenamefont {Bienias}, \citenamefont
  {Iadecola},\ and\ \citenamefont {Gorshkov}}]{PhysRevLett.128.013603}%
  \BibitemOpen
  \bibfield  {author} {\bibinfo {author} {\bibfnamefont {Fangli}\ \bibnamefont
  {Liu}}, \bibinfo {author} {\bibfnamefont {Zhi-Cheng}\ \bibnamefont {Yang}},
  \bibinfo {author} {\bibfnamefont {Przemyslaw}\ \bibnamefont {Bienias}},
  \bibinfo {author} {\bibfnamefont {Thomas}\ \bibnamefont {Iadecola}}, \ and\
  \bibinfo {author} {\bibfnamefont {Alexey~V.}\ \bibnamefont {Gorshkov}},\
  }\bibfield  {title} {\enquote {\bibinfo {title} {Localization and criticality
  in antiblockaded two-dimensional rydberg atom arrays},}\ }\href {\doibase
  10.1103/PhysRevLett.128.013603} {\bibfield  {journal} {\bibinfo  {journal}
  {Phys. Rev. Lett.}\ }\textbf {\bibinfo {volume} {128}},\ \bibinfo {pages}
  {013603} (\bibinfo {year} {2022}{\natexlab{a}})}\BibitemShut {NoStop}%
\bibitem [{\citenamefont {Kitson}\ \emph {et~al.}(2024)\citenamefont {Kitson},
  \citenamefont {Haug}, \citenamefont {La~Magna}, \citenamefont {Morsch},\ and\
  \citenamefont {Amico}}]{PhysRevA.110.043304}%
  \BibitemOpen
  \bibfield  {author} {\bibinfo {author} {\bibfnamefont {Philip}\ \bibnamefont
  {Kitson}}, \bibinfo {author} {\bibfnamefont {Tobias}\ \bibnamefont {Haug}},
  \bibinfo {author} {\bibfnamefont {Antonino}\ \bibnamefont {La~Magna}},
  \bibinfo {author} {\bibfnamefont {Oliver}\ \bibnamefont {Morsch}}, \ and\
  \bibinfo {author} {\bibfnamefont {Luigi}\ \bibnamefont {Amico}},\ }\bibfield
  {title} {\enquote {\bibinfo {title} {Rydberg atomtronic devices},}\ }\href
  {\doibase 10.1103/PhysRevA.110.043304} {\bibfield  {journal} {\bibinfo
  {journal} {Phys. Rev. A}\ }\textbf {\bibinfo {volume} {110}},\ \bibinfo
  {pages} {043304} (\bibinfo {year} {2024})}\BibitemShut {NoStop}%
\bibitem [{\citenamefont {Li}\ \emph {et~al.}(2024{\natexlab{a}})\citenamefont
  {Li}, \citenamefont {Wu}, \citenamefont {Su},\ and\ \citenamefont
  {Qian}}]{PhysRevA.109.012608}%
  \BibitemOpen
  \bibfield  {author} {\bibinfo {author} {\bibfnamefont {Wan-Xia}\ \bibnamefont
  {Li}}, \bibinfo {author} {\bibfnamefont {Jin-Lei}\ \bibnamefont {Wu}},
  \bibinfo {author} {\bibfnamefont {Shi-Lei}\ \bibnamefont {Su}}, \ and\
  \bibinfo {author} {\bibfnamefont {Jing}\ \bibnamefont {Qian}},\ }\bibfield
  {title} {\enquote {\bibinfo {title} {High-tolerance antiblockade
  $\mathrm{SWAP}$ gates using optimal pulse drivings},}\ }\href {\doibase
  10.1103/PhysRevA.109.012608} {\bibfield  {journal} {\bibinfo  {journal}
  {Phys. Rev. A}\ }\textbf {\bibinfo {volume} {109}},\ \bibinfo {pages}
  {012608} (\bibinfo {year} {2024}{\natexlab{a}})}\BibitemShut {NoStop}%
\bibitem [{\citenamefont {Jo}\ \emph {et~al.}(2020)\citenamefont {Jo},
  \citenamefont {Song}, \citenamefont {Kim},\ and\ \citenamefont
  {Ahn}}]{PhysRevLett.124.033603}%
  \BibitemOpen
  \bibfield  {author} {\bibinfo {author} {\bibfnamefont {Hanlae}\ \bibnamefont
  {Jo}}, \bibinfo {author} {\bibfnamefont {Yunheung}\ \bibnamefont {Song}},
  \bibinfo {author} {\bibfnamefont {Minhyuk}\ \bibnamefont {Kim}}, \ and\
  \bibinfo {author} {\bibfnamefont {Jaewook}\ \bibnamefont {Ahn}},\ }\bibfield
  {title} {\enquote {\bibinfo {title} {Rydberg atom entanglements in the weak
  coupling regime},}\ }\href {\doibase 10.1103/PhysRevLett.124.033603}
  {\bibfield  {journal} {\bibinfo  {journal} {Phys. Rev. Lett.}\ }\textbf
  {\bibinfo {volume} {124}},\ \bibinfo {pages} {033603} (\bibinfo {year}
  {2020})}\BibitemShut {NoStop}%
\bibitem [{\citenamefont {Wu}\ \emph {et~al.}(2020{\natexlab{a}})\citenamefont
  {Wu}, \citenamefont {Su}, \citenamefont {Wang}, \citenamefont {Song},
  \citenamefont {Xia},\ and\ \citenamefont {Jiang}}]{Wu:20}%
  \BibitemOpen
  \bibfield  {author} {\bibinfo {author} {\bibfnamefont {Jin-Lei}\ \bibnamefont
  {Wu}}, \bibinfo {author} {\bibfnamefont {Shi-Lei}\ \bibnamefont {Su}},
  \bibinfo {author} {\bibfnamefont {Yan}\ \bibnamefont {Wang}}, \bibinfo
  {author} {\bibfnamefont {Jie}\ \bibnamefont {Song}}, \bibinfo {author}
  {\bibfnamefont {Yan}\ \bibnamefont {Xia}}, \ and\ \bibinfo {author}
  {\bibfnamefont {Yong-Yuan}\ \bibnamefont {Jiang}},\ }\bibfield  {title}
  {\enquote {\bibinfo {title} {Effective rabi dynamics of rydberg atoms and
  robust high-fidelity quantum gates with a resonant amplitude-modulation
  field},}\ }\href {\doibase 10.1364/OL.386765} {\bibfield  {journal} {\bibinfo
   {journal} {Opt. Lett.}\ }\textbf {\bibinfo {volume} {45}},\ \bibinfo {pages}
  {1200--1203} (\bibinfo {year} {2020}{\natexlab{a}})}\BibitemShut {NoStop}%
\bibitem [{\citenamefont {Wu}\ \emph {et~al.}(2021{\natexlab{a}})\citenamefont
  {Wu}, \citenamefont {Wang}, \citenamefont {Han}, \citenamefont {Su},
  \citenamefont {Xia}, \citenamefont {Jiang},\ and\ \citenamefont
  {Song}}]{PhysRevA.103.012601}%
  \BibitemOpen
  \bibfield  {author} {\bibinfo {author} {\bibfnamefont {Jin-Lei}\ \bibnamefont
  {Wu}}, \bibinfo {author} {\bibfnamefont {Yan}\ \bibnamefont {Wang}}, \bibinfo
  {author} {\bibfnamefont {Jin-Xuan}\ \bibnamefont {Han}}, \bibinfo {author}
  {\bibfnamefont {Shi-Lei}\ \bibnamefont {Su}}, \bibinfo {author}
  {\bibfnamefont {Yan}\ \bibnamefont {Xia}}, \bibinfo {author} {\bibfnamefont
  {Yongyuan}\ \bibnamefont {Jiang}}, \ and\ \bibinfo {author} {\bibfnamefont
  {Jie}\ \bibnamefont {Song}},\ }\bibfield  {title} {\enquote {\bibinfo {title}
  {Resilient quantum gates on periodically driven rydberg atoms},}\ }\href
  {\doibase 10.1103/PhysRevA.103.012601} {\bibfield  {journal} {\bibinfo
  {journal} {Phys. Rev. A}\ }\textbf {\bibinfo {volume} {103}},\ \bibinfo
  {pages} {012601} (\bibinfo {year} {2021}{\natexlab{a}})}\BibitemShut
  {NoStop}%
\bibitem [{\citenamefont {Shao}\ \emph {et~al.}(2017)\citenamefont {Shao},
  \citenamefont {Wu},\ and\ \citenamefont {Yi}}]{PhysRevA.95.022317}%
  \BibitemOpen
  \bibfield  {author} {\bibinfo {author} {\bibfnamefont {Xiao-Qiang}\
  \bibnamefont {Shao}}, \bibinfo {author} {\bibfnamefont {Jin-Hui}\
  \bibnamefont {Wu}}, \ and\ \bibinfo {author} {\bibfnamefont {Xue-Xi}\
  \bibnamefont {Yi}},\ }\bibfield  {title} {\enquote {\bibinfo {title}
  {Dissipative stabilization of quantum-feedback-based multipartite
  entanglement with rydberg atoms},}\ }\href {\doibase
  10.1103/PhysRevA.95.022317} {\bibfield  {journal} {\bibinfo  {journal} {Phys.
  Rev. A}\ }\textbf {\bibinfo {volume} {95}},\ \bibinfo {pages} {022317}
  (\bibinfo {year} {2017})}\BibitemShut {NoStop}%
\bibitem [{\citenamefont {Eckardt}(2017)}]{RevModPhys.89.011004}%
  \BibitemOpen
  \bibfield  {author} {\bibinfo {author} {\bibfnamefont {Andr\'e}\ \bibnamefont
  {Eckardt}},\ }\bibfield  {title} {\enquote {\bibinfo {title} {Colloquium:
  Atomic quantum gases in periodically driven optical lattices},}\ }\href
  {\doibase 10.1103/RevModPhys.89.011004} {\bibfield  {journal} {\bibinfo
  {journal} {Rev. Mod. Phys.}\ }\textbf {\bibinfo {volume} {89}},\ \bibinfo
  {pages} {011004} (\bibinfo {year} {2017})}\BibitemShut {NoStop}%
\bibitem [{\citenamefont {Basak}\ \emph {et~al.}(2018)\citenamefont {Basak},
  \citenamefont {Chougale},\ and\ \citenamefont
  {Nath}}]{PhysRevLett.120.123204}%
  \BibitemOpen
  \bibfield  {author} {\bibinfo {author} {\bibfnamefont {Sagarika}\
  \bibnamefont {Basak}}, \bibinfo {author} {\bibfnamefont {Yashwant}\
  \bibnamefont {Chougale}}, \ and\ \bibinfo {author} {\bibfnamefont {Rejish}\
  \bibnamefont {Nath}},\ }\bibfield  {title} {\enquote {\bibinfo {title}
  {Periodically driven array of single rydberg atoms},}\ }\href {\doibase
  10.1103/PhysRevLett.120.123204} {\bibfield  {journal} {\bibinfo  {journal}
  {Phys. Rev. Lett.}\ }\textbf {\bibinfo {volume} {120}},\ \bibinfo {pages}
  {123204} (\bibinfo {year} {2018})}\BibitemShut {NoStop}%
\bibitem [{\citenamefont {Mallavarapu}\ \emph {et~al.}(2021)\citenamefont
  {Mallavarapu}, \citenamefont {Niranjan}, \citenamefont {Li}, \citenamefont
  {W\"uster},\ and\ \citenamefont {Nath}}]{PhysRevA.103.023335}%
  \BibitemOpen
  \bibfield  {author} {\bibinfo {author} {\bibfnamefont {S.~Kumar}\
  \bibnamefont {Mallavarapu}}, \bibinfo {author} {\bibfnamefont {Ankita}\
  \bibnamefont {Niranjan}}, \bibinfo {author} {\bibfnamefont {Weibin}\
  \bibnamefont {Li}}, \bibinfo {author} {\bibfnamefont {Sebastian}\
  \bibnamefont {W\"uster}}, \ and\ \bibinfo {author} {\bibfnamefont {Rejish}\
  \bibnamefont {Nath}},\ }\bibfield  {title} {\enquote {\bibinfo {title}
  {Population trapping in a pair of periodically driven rydberg atoms},}\
  }\href {\doibase 10.1103/PhysRevA.103.023335} {\bibfield  {journal} {\bibinfo
   {journal} {Phys. Rev. A}\ }\textbf {\bibinfo {volume} {103}},\ \bibinfo
  {pages} {023335} (\bibinfo {year} {2021})}\BibitemShut {NoStop}%
\bibitem [{\citenamefont {Nguyen}\ \emph {et~al.}(2024)\citenamefont {Nguyen},
  \citenamefont {Kim}, \citenamefont {Hashim}, \citenamefont {Goss},
  \citenamefont {Marinelli}, \citenamefont {Bhandari}, \citenamefont {Das},
  \citenamefont {Naik}, \citenamefont {Kreikebaum}, \citenamefont {Jordan},
  \citenamefont {Santiago},\ and\ \citenamefont {Siddiqi}}]{Nguyen2024}%
  \BibitemOpen
  \bibfield  {author} {\bibinfo {author} {\bibfnamefont {Long~B.}\ \bibnamefont
  {Nguyen}}, \bibinfo {author} {\bibfnamefont {Yosep}\ \bibnamefont {Kim}},
  \bibinfo {author} {\bibfnamefont {Akel}\ \bibnamefont {Hashim}}, \bibinfo
  {author} {\bibfnamefont {Noah}\ \bibnamefont {Goss}}, \bibinfo {author}
  {\bibfnamefont {Brian}\ \bibnamefont {Marinelli}}, \bibinfo {author}
  {\bibfnamefont {Bibek}\ \bibnamefont {Bhandari}}, \bibinfo {author}
  {\bibfnamefont {Debmalya}\ \bibnamefont {Das}}, \bibinfo {author}
  {\bibfnamefont {Ravi~K.}\ \bibnamefont {Naik}}, \bibinfo {author}
  {\bibfnamefont {John~Mark}\ \bibnamefont {Kreikebaum}}, \bibinfo {author}
  {\bibfnamefont {Andrew~N.}\ \bibnamefont {Jordan}}, \bibinfo {author}
  {\bibfnamefont {David~I.}\ \bibnamefont {Santiago}}, \ and\ \bibinfo {author}
  {\bibfnamefont {Irfan}\ \bibnamefont {Siddiqi}},\ }\bibfield  {title}
  {\enquote {\bibinfo {title} {Programmable heisenberg interactions between
  floquet qubits},}\ }\href {\doibase 10.1038/s41567-023-02326-7} {\bibfield
  {journal} {\bibinfo  {journal} {Nature Physics}\ }\textbf {\bibinfo {volume}
  {20}},\ \bibinfo {pages} {240--246} (\bibinfo {year} {2024})}\BibitemShut
  {NoStop}%
\bibitem [{\citenamefont {Zhou}\ \emph {et~al.}(2024)\citenamefont {Zhou},
  \citenamefont {Liu}, \citenamefont {Liu}, \citenamefont {Lu}, \citenamefont
  {Li}, \citenamefont {Xie}, \citenamefont {Lydick}, \citenamefont {Hao},
  \citenamefont {Liu}, \citenamefont {Watanabe}, \citenamefont {Taniguchi},
  \citenamefont {Chou}, \citenamefont {Forrest},\ and\ \citenamefont
  {Deng}}]{Zhou2024}%
  \BibitemOpen
  \bibfield  {author} {\bibinfo {author} {\bibfnamefont {Lingxiao}\
  \bibnamefont {Zhou}}, \bibinfo {author} {\bibfnamefont {Bin}\ \bibnamefont
  {Liu}}, \bibinfo {author} {\bibfnamefont {Yuze}\ \bibnamefont {Liu}},
  \bibinfo {author} {\bibfnamefont {Yang}\ \bibnamefont {Lu}}, \bibinfo
  {author} {\bibfnamefont {Qiuyang}\ \bibnamefont {Li}}, \bibinfo {author}
  {\bibfnamefont {Xin}\ \bibnamefont {Xie}}, \bibinfo {author} {\bibfnamefont
  {Nathanial}\ \bibnamefont {Lydick}}, \bibinfo {author} {\bibfnamefont
  {Ruofan}\ \bibnamefont {Hao}}, \bibinfo {author} {\bibfnamefont {Chenxi}\
  \bibnamefont {Liu}}, \bibinfo {author} {\bibfnamefont {Kenji}\ \bibnamefont
  {Watanabe}}, \bibinfo {author} {\bibfnamefont {Takashi}\ \bibnamefont
  {Taniguchi}}, \bibinfo {author} {\bibfnamefont {Yu-Hsun}\ \bibnamefont
  {Chou}}, \bibinfo {author} {\bibfnamefont {Stephen~R.}\ \bibnamefont
  {Forrest}}, \ and\ \bibinfo {author} {\bibfnamefont {Hui}\ \bibnamefont
  {Deng}},\ }\bibfield  {title} {\enquote {\bibinfo {title} {Cavity floquet
  engineering},}\ }\href {\doibase 10.1038/s41467-024-52014-0} {\bibfield
  {journal} {\bibinfo  {journal} {Nature Communications}\ }\textbf {\bibinfo
  {volume} {15}},\ \bibinfo {pages} {7782} (\bibinfo {year}
  {2024})}\BibitemShut {NoStop}%
\bibitem [{\citenamefont {Zhao}\ \emph {et~al.}(2023)\citenamefont {Zhao},
  \citenamefont {Lee}, \citenamefont {Aliyu},\ and\ \citenamefont
  {Loh}}]{Zhao2023}%
  \BibitemOpen
  \bibfield  {author} {\bibinfo {author} {\bibfnamefont {Luheng}\ \bibnamefont
  {Zhao}}, \bibinfo {author} {\bibfnamefont {Michael Dao~Kang}\ \bibnamefont
  {Lee}}, \bibinfo {author} {\bibfnamefont {Mohammad~Mujahid}\ \bibnamefont
  {Aliyu}}, \ and\ \bibinfo {author} {\bibfnamefont {Huanqian}\ \bibnamefont
  {Loh}},\ }\bibfield  {title} {\enquote {\bibinfo {title} {Floquet-tailored
  rydberg interactions},}\ }\href {\doibase 10.1038/s41467-023-42899-8}
  {\bibfield  {journal} {\bibinfo  {journal} {Nature Communications}\ }\textbf
  {\bibinfo {volume} {14}},\ \bibinfo {pages} {7128} (\bibinfo {year}
  {2023})}\BibitemShut {NoStop}%
\bibitem [{\citenamefont {Heya}\ \emph {et~al.}(2024)\citenamefont {Heya},
  \citenamefont {Malekakhlagh}, \citenamefont {Merkel}, \citenamefont
  {Kanazawa},\ and\ \citenamefont {Pritchett}}]{PhysRevApplied.21.024035}%
  \BibitemOpen
  \bibfield  {author} {\bibinfo {author} {\bibfnamefont {Kentaro}\ \bibnamefont
  {Heya}}, \bibinfo {author} {\bibfnamefont {Moein}\ \bibnamefont
  {Malekakhlagh}}, \bibinfo {author} {\bibfnamefont {Seth}\ \bibnamefont
  {Merkel}}, \bibinfo {author} {\bibfnamefont {Naoki}\ \bibnamefont
  {Kanazawa}}, \ and\ \bibinfo {author} {\bibfnamefont {Emily}\ \bibnamefont
  {Pritchett}},\ }\bibfield  {title} {\enquote {\bibinfo {title} {Floquet
  analysis of frequency collisions},}\ }\href {\doibase
  10.1103/PhysRevApplied.21.024035} {\bibfield  {journal} {\bibinfo  {journal}
  {Phys. Rev. Appl.}\ }\textbf {\bibinfo {volume} {21}},\ \bibinfo {pages}
  {024035} (\bibinfo {year} {2024})}\BibitemShut {NoStop}%
\bibitem [{\citenamefont {Sun}\ \emph {et~al.}(2024)\citenamefont {Sun},
  \citenamefont {Wu},\ and\ \citenamefont {Su}}]{Sun_2024}%
  \BibitemOpen
  \bibfield  {author} {\bibinfo {author} {\bibfnamefont {Hao-Wen}\ \bibnamefont
  {Sun}}, \bibinfo {author} {\bibfnamefont {Jin-Lei}\ \bibnamefont {Wu}}, \
  and\ \bibinfo {author} {\bibfnamefont {Shi-Lei}\ \bibnamefont {Su}},\
  }\bibfield  {title} {\enquote {\bibinfo {title} {Floquet geometric entangling
  gates in ground-state manifolds of rydberg atoms},}\ }\href {\doibase
  10.1088/1402-4896/ad635b} {\bibfield  {journal} {\bibinfo  {journal} {Physica
  Scripta}\ }\textbf {\bibinfo {volume} {99}},\ \bibinfo {pages} {085122}
  (\bibinfo {year} {2024})}\BibitemShut {NoStop}%
\bibitem [{\citenamefont {Liu}\ \emph {et~al.}(2019)\citenamefont {Liu},
  \citenamefont {Song}, \citenamefont {Xue}, \citenamefont {Wang},\ and\
  \citenamefont {Yung}}]{PhysRevLett.123.100501}%
  \BibitemOpen
  \bibfield  {author} {\bibinfo {author} {\bibfnamefont {Bao-Jie}\ \bibnamefont
  {Liu}}, \bibinfo {author} {\bibfnamefont {Xue-Ke}\ \bibnamefont {Song}},
  \bibinfo {author} {\bibfnamefont {Zheng-Yuan}\ \bibnamefont {Xue}}, \bibinfo
  {author} {\bibfnamefont {Xin}\ \bibnamefont {Wang}}, \ and\ \bibinfo {author}
  {\bibfnamefont {Man-Hong}\ \bibnamefont {Yung}},\ }\bibfield  {title}
  {\enquote {\bibinfo {title} {Plug-and-play approach to nonadiabatic geometric
  quantum gates},}\ }\href {\doibase 10.1103/PhysRevLett.123.100501} {\bibfield
   {journal} {\bibinfo  {journal} {Phys. Rev. Lett.}\ }\textbf {\bibinfo
  {volume} {123}},\ \bibinfo {pages} {100501} (\bibinfo {year}
  {2019})}\BibitemShut {NoStop}%
\bibitem [{\citenamefont {Li}\ \emph {et~al.}(2017)\citenamefont {Li},
  \citenamefont {Liu},\ and\ \citenamefont {Long}}]{Li2017}%
  \BibitemOpen
  \bibfield  {author} {\bibinfo {author} {\bibfnamefont {Hang}\ \bibnamefont
  {Li}}, \bibinfo {author} {\bibfnamefont {Yang}\ \bibnamefont {Liu}}, \ and\
  \bibinfo {author} {\bibfnamefont {GuiLu}\ \bibnamefont {Long}},\ }\bibfield
  {title} {\enquote {\bibinfo {title} {Experimental realization of single-shot
  nonadiabatic holonomic gates in nuclear spins},}\ }\href {\doibase
  10.1007/s11433-017-9058-7} {\bibfield  {journal} {\bibinfo  {journal}
  {Science China Physics, Mechanics {\&} Astronomy}\ }\textbf {\bibinfo
  {volume} {60}},\ \bibinfo {pages} {080311} (\bibinfo {year}
  {2017})}\BibitemShut {NoStop}%
\bibitem [{\citenamefont {Ma}\ \emph {et~al.}(2023)\citenamefont {Ma},
  \citenamefont {Liu}, \citenamefont {Peng}, \citenamefont {Zhang},
  \citenamefont {Jandura}, \citenamefont {Claes}, \citenamefont {Burgers},
  \citenamefont {Pupillo}, \citenamefont {Puri},\ and\ \citenamefont
  {Thompson}}]{Ma2023}%
  \BibitemOpen
  \bibfield  {author} {\bibinfo {author} {\bibfnamefont {Shuo}\ \bibnamefont
  {Ma}}, \bibinfo {author} {\bibfnamefont {Genyue}\ \bibnamefont {Liu}},
  \bibinfo {author} {\bibfnamefont {Pai}\ \bibnamefont {Peng}}, \bibinfo
  {author} {\bibfnamefont {Bichen}\ \bibnamefont {Zhang}}, \bibinfo {author}
  {\bibfnamefont {Sven}\ \bibnamefont {Jandura}}, \bibinfo {author}
  {\bibfnamefont {Jahan}\ \bibnamefont {Claes}}, \bibinfo {author}
  {\bibfnamefont {Alex~P.}\ \bibnamefont {Burgers}}, \bibinfo {author}
  {\bibfnamefont {Guido}\ \bibnamefont {Pupillo}}, \bibinfo {author}
  {\bibfnamefont {Shruti}\ \bibnamefont {Puri}}, \ and\ \bibinfo {author}
  {\bibfnamefont {Jeff~D.}\ \bibnamefont {Thompson}},\ }\bibfield  {title}
  {\enquote {\bibinfo {title} {High-fidelity gates and mid-circuit erasure
  conversion in an atomic qubit},}\ }\href {\doibase
  10.1038/s41586-023-06438-1} {\bibfield  {journal} {\bibinfo  {journal}
  {Nature}\ }\textbf {\bibinfo {volume} {622}},\ \bibinfo {pages} {279--284}
  (\bibinfo {year} {2023})}\BibitemShut {NoStop}%
\bibitem [{\citenamefont {Song}\ \emph
  {et~al.}(2016{\natexlab{a}})\citenamefont {Song}, \citenamefont {Zhang},
  \citenamefont {Ai}, \citenamefont {Qiu},\ and\ \citenamefont
  {Deng}}]{Song_2016}%
  \BibitemOpen
  \bibfield  {author} {\bibinfo {author} {\bibfnamefont {Xue-Ke}\ \bibnamefont
  {Song}}, \bibinfo {author} {\bibfnamefont {Hao}\ \bibnamefont {Zhang}},
  \bibinfo {author} {\bibfnamefont {Qing}\ \bibnamefont {Ai}}, \bibinfo
  {author} {\bibfnamefont {Jing}\ \bibnamefont {Qiu}}, \ and\ \bibinfo {author}
  {\bibfnamefont {Fu-Guo}\ \bibnamefont {Deng}},\ }\bibfield  {title} {\enquote
  {\bibinfo {title} {Shortcuts to adiabatic holonomic quantum computation in
  decoherence-free subspace with transitionless quantum driving algorithm},}\
  }\href {\doibase 10.1088/1367-2630/18/2/023001} {\bibfield  {journal}
  {\bibinfo  {journal} {New Journal of Physics}\ }\textbf {\bibinfo {volume}
  {18}},\ \bibinfo {pages} {023001} (\bibinfo {year}
  {2016}{\natexlab{a}})}\BibitemShut {NoStop}%
\bibitem [{\citenamefont {Zhang}\ \emph {et~al.}(2025)\citenamefont {Zhang},
  \citenamefont {Wei}, \citenamefont {Song}, \citenamefont {Yan}, \citenamefont
  {Kinos}, \citenamefont {Su},\ and\ \citenamefont
  {Chen}}]{PhysRevA.111.012604}%
  \BibitemOpen
  \bibfield  {author} {\bibinfo {author} {\bibfnamefont {S.-Y.}\ \bibnamefont
  {Zhang}}, \bibinfo {author} {\bibfnamefont {J.-F.}\ \bibnamefont {Wei}},
  \bibinfo {author} {\bibfnamefont {P.-Y.}\ \bibnamefont {Song}}, \bibinfo
  {author} {\bibfnamefont {L.-L.}\ \bibnamefont {Yan}}, \bibinfo {author}
  {\bibfnamefont {A.}~\bibnamefont {Kinos}}, \bibinfo {author} {\bibfnamefont
  {Shi-Lei}\ \bibnamefont {Su}}, \ and\ \bibinfo {author} {\bibfnamefont
  {Gang}\ \bibnamefont {Chen}},\ }\bibfield  {title} {\enquote {\bibinfo
  {title} {Quantum computation based on capture-and-release dynamics},}\ }\href
  {\doibase 10.1103/PhysRevA.111.012604} {\bibfield  {journal} {\bibinfo
  {journal} {Phys. Rev. A}\ }\textbf {\bibinfo {volume} {111}},\ \bibinfo
  {pages} {012604} (\bibinfo {year} {2025})}\BibitemShut {NoStop}%
\bibitem [{\citenamefont {Haase}\ \emph {et~al.}(2018)\citenamefont {Haase},
  \citenamefont {Wang}, \citenamefont {Casanova},\ and\ \citenamefont
  {Plenio}}]{PhysRevLett.121.050402}%
  \BibitemOpen
  \bibfield  {author} {\bibinfo {author} {\bibfnamefont {J.~F.}\ \bibnamefont
  {Haase}}, \bibinfo {author} {\bibfnamefont {Z.-Y.}\ \bibnamefont {Wang}},
  \bibinfo {author} {\bibfnamefont {J.}~\bibnamefont {Casanova}}, \ and\
  \bibinfo {author} {\bibfnamefont {M.~B.}\ \bibnamefont {Plenio}},\ }\bibfield
   {title} {\enquote {\bibinfo {title} {Soft quantum control for highly
  selective interactions among joint quantum systems},}\ }\href {\doibase
  10.1103/PhysRevLett.121.050402} {\bibfield  {journal} {\bibinfo  {journal}
  {Phys. Rev. Lett.}\ }\textbf {\bibinfo {volume} {121}},\ \bibinfo {pages}
  {050402} (\bibinfo {year} {2018})}\BibitemShut {NoStop}%
\bibitem [{\citenamefont {Wu}\ \emph {et~al.}(2021{\natexlab{b}})\citenamefont
  {Wu}, \citenamefont {Tang}, \citenamefont {Wang}, \citenamefont {Wang},
  \citenamefont {Han}, \citenamefont {L{\"u}}, \citenamefont {Song},
  \citenamefont {Su}, \citenamefont {Xia},\ and\ \citenamefont
  {Jiang}}]{Wu2021}%
  \BibitemOpen
  \bibfield  {author} {\bibinfo {author} {\bibfnamefont {JinLei}\ \bibnamefont
  {Wu}}, \bibinfo {author} {\bibfnamefont {Shuai}\ \bibnamefont {Tang}},
  \bibinfo {author} {\bibfnamefont {Yan}\ \bibnamefont {Wang}}, \bibinfo
  {author} {\bibfnamefont {XiaoSai}\ \bibnamefont {Wang}}, \bibinfo {author}
  {\bibfnamefont {JinXuan}\ \bibnamefont {Han}}, \bibinfo {author}
  {\bibfnamefont {Cheng}\ \bibnamefont {L{\"u}}}, \bibinfo {author}
  {\bibfnamefont {Jie}\ \bibnamefont {Song}}, \bibinfo {author} {\bibfnamefont
  {ShiLei}\ \bibnamefont {Su}}, \bibinfo {author} {\bibfnamefont {Yan}\
  \bibnamefont {Xia}}, \ and\ \bibinfo {author} {\bibfnamefont {YongYuan}\
  \bibnamefont {Jiang}},\ }\bibfield  {title} {\enquote {\bibinfo {title}
  {Unidirectional acoustic metamaterials based on nonadiabatic holonomic
  quantum transformations},}\ }\href {\doibase 10.1007/s11433-021-1810-6}
  {\bibfield  {journal} {\bibinfo  {journal} {Science China Physics, Mechanics
  {\&} Astronomy}\ }\textbf {\bibinfo {volume} {65}},\ \bibinfo {pages}
  {220311} (\bibinfo {year} {2021}{\natexlab{b}})}\BibitemShut {NoStop}%
\bibitem [{\citenamefont {Yin}\ and\ \citenamefont
  {Shao}(2021)}]{Yin2021GaussianSC}%
  \BibitemOpen
  \bibfield  {author} {\bibinfo {author} {\bibfnamefont {Hong-Da}\ \bibnamefont
  {Yin}}\ and\ \bibinfo {author} {\bibfnamefont {X.~Q.}\ \bibnamefont {Shao}},\
  }\bibfield  {title} {\enquote {\bibinfo {title} {Gaussian soft control-based
  quantum fan-out gate in ground-state manifolds of neutral atoms.}}\ }\href
  {https://api.semanticscholar.org/CorpusID:23449715} {\bibfield  {journal}
  {\bibinfo  {journal} {Optics letters}\ }\textbf {\bibinfo {volume} {46 10}},\
  \bibinfo {pages} {2541--2544} (\bibinfo {year} {2021})}\BibitemShut {NoStop}%
\bibitem [{\citenamefont {Song}\ \emph
  {et~al.}(2016{\natexlab{b}})\citenamefont {Song}, \citenamefont {Ai},
  \citenamefont {Qiu},\ and\ \citenamefont {Deng}}]{PhysRevA.93.052324}%
  \BibitemOpen
  \bibfield  {author} {\bibinfo {author} {\bibfnamefont {Xue-Ke}\ \bibnamefont
  {Song}}, \bibinfo {author} {\bibfnamefont {Qing}\ \bibnamefont {Ai}},
  \bibinfo {author} {\bibfnamefont {Jing}\ \bibnamefont {Qiu}}, \ and\ \bibinfo
  {author} {\bibfnamefont {Fu-Guo}\ \bibnamefont {Deng}},\ }\bibfield  {title}
  {\enquote {\bibinfo {title} {Physically feasible three-level transitionless
  quantum driving with multiple schr\"odinger dynamics},}\ }\href {\doibase
  10.1103/PhysRevA.93.052324} {\bibfield  {journal} {\bibinfo  {journal} {Phys.
  Rev. A}\ }\textbf {\bibinfo {volume} {93}},\ \bibinfo {pages} {052324}
  (\bibinfo {year} {2016}{\natexlab{b}})}\BibitemShut {NoStop}%
\bibitem [{\citenamefont {Wu}\ \emph {et~al.}(2020{\natexlab{b}})\citenamefont
  {Wu}, \citenamefont {Wang}, \citenamefont {Han}, \citenamefont {Wang},
  \citenamefont {Su}, \citenamefont {Xia}, \citenamefont {Jiang},\ and\
  \citenamefont {Song}}]{PhysRevApplied.13.044021}%
  \BibitemOpen
  \bibfield  {author} {\bibinfo {author} {\bibfnamefont {Jin-Lei}\ \bibnamefont
  {Wu}}, \bibinfo {author} {\bibfnamefont {Yan}\ \bibnamefont {Wang}}, \bibinfo
  {author} {\bibfnamefont {Jin-Xuan}\ \bibnamefont {Han}}, \bibinfo {author}
  {\bibfnamefont {Cong}\ \bibnamefont {Wang}}, \bibinfo {author} {\bibfnamefont
  {Shi-Lei}\ \bibnamefont {Su}}, \bibinfo {author} {\bibfnamefont {Yan}\
  \bibnamefont {Xia}}, \bibinfo {author} {\bibfnamefont {Yongyuan}\
  \bibnamefont {Jiang}}, \ and\ \bibinfo {author} {\bibfnamefont {Jie}\
  \bibnamefont {Song}},\ }\bibfield  {title} {\enquote {\bibinfo {title}
  {Two-path interference for enantiomer-selective state transfer of chiral
  molecules},}\ }\href {\doibase 10.1103/PhysRevApplied.13.044021} {\bibfield
  {journal} {\bibinfo  {journal} {Phys. Rev. Appl.}\ }\textbf {\bibinfo
  {volume} {13}},\ \bibinfo {pages} {044021} (\bibinfo {year}
  {2020}{\natexlab{b}})}\BibitemShut {NoStop}%
\bibitem [{\citenamefont {Han}\ \emph {et~al.}(2021)\citenamefont {Han},
  \citenamefont {Wu}, \citenamefont {Wang}, \citenamefont {Xia}, \citenamefont
  {Jiang},\ and\ \citenamefont {Song}}]{PhysRevA.103.032402}%
  \BibitemOpen
  \bibfield  {author} {\bibinfo {author} {\bibfnamefont {Jin-Xuan}\
  \bibnamefont {Han}}, \bibinfo {author} {\bibfnamefont {Jin-Lei}\ \bibnamefont
  {Wu}}, \bibinfo {author} {\bibfnamefont {Yan}\ \bibnamefont {Wang}}, \bibinfo
  {author} {\bibfnamefont {Yan}\ \bibnamefont {Xia}}, \bibinfo {author}
  {\bibfnamefont {Yong-Yuan}\ \bibnamefont {Jiang}}, \ and\ \bibinfo {author}
  {\bibfnamefont {Jie}\ \bibnamefont {Song}},\ }\bibfield  {title} {\enquote
  {\bibinfo {title} {Large-scale greenberger-horne-zeilinger states through a
  topologically protected zero-energy mode in a superconducting
  qutrit-resonator chain},}\ }\href {\doibase 10.1103/PhysRevA.103.032402}
  {\bibfield  {journal} {\bibinfo  {journal} {Phys. Rev. A}\ }\textbf {\bibinfo
  {volume} {103}},\ \bibinfo {pages} {032402} (\bibinfo {year}
  {2021})}\BibitemShut {NoStop}%
\bibitem [{\citenamefont {Wu}\ \emph {et~al.}(2024)\citenamefont {Wu},
  \citenamefont {Xing},\ and\ \citenamefont {Yin}}]{Wu2024}%
  \BibitemOpen
  \bibfield  {author} {\bibinfo {author} {\bibfnamefont {Qiaolin}\ \bibnamefont
  {Wu}}, \bibinfo {author} {\bibfnamefont {Jun}\ \bibnamefont {Xing}}, \ and\
  \bibinfo {author} {\bibfnamefont {Hongda}\ \bibnamefont {Yin}},\ }\bibfield
  {title} {\enquote {\bibinfo {title} {Soft-controlled quantum gate with
  enhanced robustness and undegraded dynamics in rydberg atoms},}\ }\href
  {\doibase 10.1140/epjqt/s40507-023-00211-z} {\bibfield  {journal} {\bibinfo
  {journal} {EPJ Quantum Technology}\ }\textbf {\bibinfo {volume} {11}},\
  \bibinfo {pages} {1} (\bibinfo {year} {2024})}\BibitemShut {NoStop}%
\bibitem [{\citenamefont {Guo}\ \emph {et~al.}(2024)\citenamefont {Guo},
  \citenamefont {Wu}, \citenamefont {Cao}, \citenamefont {Zhang},\ and\
  \citenamefont {Su}}]{PhysRevA.110.043510}%
  \BibitemOpen
  \bibfield  {author} {\bibinfo {author} {\bibfnamefont {Jin-Kang}\
  \bibnamefont {Guo}}, \bibinfo {author} {\bibfnamefont {Jin-Lei}\ \bibnamefont
  {Wu}}, \bibinfo {author} {\bibfnamefont {Ji}~\bibnamefont {Cao}}, \bibinfo
  {author} {\bibfnamefont {Shou}\ \bibnamefont {Zhang}}, \ and\ \bibinfo
  {author} {\bibfnamefont {Shi-Lei}\ \bibnamefont {Su}},\ }\bibfield  {title}
  {\enquote {\bibinfo {title} {Shortcut engineering for accelerating
  topological quantum state transfers in optomechanical lattices},}\ }\href
  {\doibase 10.1103/PhysRevA.110.043510} {\bibfield  {journal} {\bibinfo
  {journal} {Phys. Rev. A}\ }\textbf {\bibinfo {volume} {110}},\ \bibinfo
  {pages} {043510} (\bibinfo {year} {2024})}\BibitemShut {NoStop}%
\bibitem [{\citenamefont {Guo}\ \emph {et~al.}(2025)\citenamefont {Guo},
  \citenamefont {Su}, \citenamefont {Li},\ and\ \citenamefont
  {Shao}}]{PhysRevA.111.022420}%
  \BibitemOpen
  \bibfield  {author} {\bibinfo {author} {\bibfnamefont {F.~Q.}\ \bibnamefont
  {Guo}}, \bibinfo {author} {\bibfnamefont {Shi-Lei}\ \bibnamefont {Su}},
  \bibinfo {author} {\bibfnamefont {Weibin}\ \bibnamefont {Li}}, \ and\
  \bibinfo {author} {\bibfnamefont {X.~Q.}\ \bibnamefont {Shao}},\ }\bibfield
  {title} {\enquote {\bibinfo {title} {Parity-controlled gate in a
  two-dimensional neutral-atom array},}\ }\href {\doibase
  10.1103/PhysRevA.111.022420} {\bibfield  {journal} {\bibinfo  {journal}
  {Phys. Rev. A}\ }\textbf {\bibinfo {volume} {111}},\ \bibinfo {pages}
  {022420} (\bibinfo {year} {2025})}\BibitemShut {NoStop}%
\bibitem [{\citenamefont {Grover}(1997)}]{PhysRevLett.79.325}%
  \BibitemOpen
  \bibfield  {author} {\bibinfo {author} {\bibfnamefont {Lov~K.}\ \bibnamefont
  {Grover}},\ }\bibfield  {title} {\enquote {\bibinfo {title} {Quantum
  mechanics helps in searching for a needle in a haystack},}\ }\href {\doibase
  10.1103/PhysRevLett.79.325} {\bibfield  {journal} {\bibinfo  {journal} {Phys.
  Rev. Lett.}\ }\textbf {\bibinfo {volume} {79}},\ \bibinfo {pages} {325--328}
  (\bibinfo {year} {1997})}\BibitemShut {NoStop}%
\bibitem [{\citenamefont {Grover}(1998)}]{PhysRevLett.80.4329}%
  \BibitemOpen
  \bibfield  {author} {\bibinfo {author} {\bibfnamefont {Lov~K.}\ \bibnamefont
  {Grover}},\ }\bibfield  {title} {\enquote {\bibinfo {title} {Quantum
  computers can search rapidly by using almost any transformation},}\ }\href
  {\doibase 10.1103/PhysRevLett.80.4329} {\bibfield  {journal} {\bibinfo
  {journal} {Phys. Rev. Lett.}\ }\textbf {\bibinfo {volume} {80}},\ \bibinfo
  {pages} {4329--4332} (\bibinfo {year} {1998})}\BibitemShut {NoStop}%
\bibitem [{\citenamefont {Long}(2001)}]{PhysRevA.64.022307}%
  \BibitemOpen
  \bibfield  {author} {\bibinfo {author} {\bibfnamefont {G.~L.}\ \bibnamefont
  {Long}},\ }\bibfield  {title} {\enquote {\bibinfo {title} {Grover algorithm
  with zero theoretical failure rate},}\ }\href {\doibase
  10.1103/PhysRevA.64.022307} {\bibfield  {journal} {\bibinfo  {journal} {Phys.
  Rev. A}\ }\textbf {\bibinfo {volume} {64}},\ \bibinfo {pages} {022307}
  (\bibinfo {year} {2001})}\BibitemShut {NoStop}%
\bibitem [{\citenamefont {Long}\ \emph {et~al.}(1999)\citenamefont {Long},
  \citenamefont {Li}, \citenamefont {Zhang},\ and\ \citenamefont
  {Niu}}]{LONG199927}%
  \BibitemOpen
  \bibfield  {author} {\bibinfo {author} {\bibfnamefont {Gui~Lu}\ \bibnamefont
  {Long}}, \bibinfo {author} {\bibfnamefont {Yan~Song}\ \bibnamefont {Li}},
  \bibinfo {author} {\bibfnamefont {Wei~Lin}\ \bibnamefont {Zhang}}, \ and\
  \bibinfo {author} {\bibfnamefont {Li}~\bibnamefont {Niu}},\ }\bibfield
  {title} {\enquote {\bibinfo {title} {Phase matching in quantum searching},}\
  }\href {\doibase https://doi.org/10.1016/S0375-9601(99)00631-3} {\bibfield
  {journal} {\bibinfo  {journal} {Physics Letters A}\ }\textbf {\bibinfo
  {volume} {262}},\ \bibinfo {pages} {27--34} (\bibinfo {year}
  {1999})}\BibitemShut {NoStop}%
\bibitem [{\citenamefont {Long}\ \emph {et~al.}(2002)\citenamefont {Long},
  \citenamefont {Li},\ and\ \citenamefont {Sun}}]{LONG2002143}%
  \BibitemOpen
  \bibfield  {author} {\bibinfo {author} {\bibfnamefont {Gui-Lu}\ \bibnamefont
  {Long}}, \bibinfo {author} {\bibfnamefont {Xiao}\ \bibnamefont {Li}}, \ and\
  \bibinfo {author} {\bibfnamefont {Yang}\ \bibnamefont {Sun}},\ }\bibfield
  {title} {\enquote {\bibinfo {title} {Phase matching condition for quantum
  search with a generalized initial state},}\ }\href {\doibase
  https://doi.org/10.1016/S0375-9601(02)00055-5} {\bibfield  {journal}
  {\bibinfo  {journal} {Physics Letters A}\ }\textbf {\bibinfo {volume}
  {294}},\ \bibinfo {pages} {143--152} (\bibinfo {year} {2002})}\BibitemShut
  {NoStop}%
\bibitem [{\citenamefont {He}\ \emph {et~al.}(2023)\citenamefont {He},
  \citenamefont {Zhao}, \citenamefont {Lv}, \citenamefont {Peng}, \citenamefont
  {Sun}, \citenamefont {Sun}, \citenamefont {Su},\ and\ \citenamefont
  {Yang}}]{He:23}%
  \BibitemOpen
  \bibfield  {author} {\bibinfo {author} {\bibfnamefont {Xin}\ \bibnamefont
  {He}}, \bibinfo {author} {\bibfnamefont {Wen-Tao}\ \bibnamefont {Zhao}},
  \bibinfo {author} {\bibfnamefont {Wang-Chu}\ \bibnamefont {Lv}}, \bibinfo
  {author} {\bibfnamefont {Chen-Hui}\ \bibnamefont {Peng}}, \bibinfo {author}
  {\bibfnamefont {Zhe}\ \bibnamefont {Sun}}, \bibinfo {author} {\bibfnamefont
  {Yong-Nan}\ \bibnamefont {Sun}}, \bibinfo {author} {\bibfnamefont {Qi-Ping}\
  \bibnamefont {Su}}, \ and\ \bibinfo {author} {\bibfnamefont {Chui-Ping}\
  \bibnamefont {Yang}},\ }\bibfield  {title} {\enquote {\bibinfo {title}
  {Experimental demonstration of deterministic quantum search for multiple
  marked states without adjusting the oracle},}\ }\href {\doibase
  10.1364/OL.497599} {\bibfield  {journal} {\bibinfo  {journal} {Opt. Lett.}\
  }\textbf {\bibinfo {volume} {48}},\ \bibinfo {pages} {4428--4431} (\bibinfo
  {year} {2023})}\BibitemShut {NoStop}%
\bibitem [{\citenamefont {Sankar}\ \emph {et~al.}(2024)\citenamefont {Sankar},
  \citenamefont {Scherer}, \citenamefont {Kako}, \citenamefont {Reifenstein},
  \citenamefont {Ghadermarzy}, \citenamefont {Krayenhoff}, \citenamefont
  {Inui}, \citenamefont {Ng}, \citenamefont {Onodera}, \citenamefont {Ronagh},\
  and\ \citenamefont {Yamamoto}}]{Sankar2024}%
  \BibitemOpen
  \bibfield  {author} {\bibinfo {author} {\bibfnamefont {Krishanu}\
  \bibnamefont {Sankar}}, \bibinfo {author} {\bibfnamefont {Artur}\
  \bibnamefont {Scherer}}, \bibinfo {author} {\bibfnamefont {Satoshi}\
  \bibnamefont {Kako}}, \bibinfo {author} {\bibfnamefont {Sam}\ \bibnamefont
  {Reifenstein}}, \bibinfo {author} {\bibfnamefont {Navid}\ \bibnamefont
  {Ghadermarzy}}, \bibinfo {author} {\bibfnamefont {Willem~B.}\ \bibnamefont
  {Krayenhoff}}, \bibinfo {author} {\bibfnamefont {Yoshitaka}\ \bibnamefont
  {Inui}}, \bibinfo {author} {\bibfnamefont {Edwin}\ \bibnamefont {Ng}},
  \bibinfo {author} {\bibfnamefont {Tatsuhiro}\ \bibnamefont {Onodera}},
  \bibinfo {author} {\bibfnamefont {Pooya}\ \bibnamefont {Ronagh}}, \ and\
  \bibinfo {author} {\bibfnamefont {Yoshihisa}\ \bibnamefont {Yamamoto}},\
  }\bibfield  {title} {\enquote {\bibinfo {title} {A benchmarking study of
  quantum algorithms for combinatorial optimization},}\ }\href {\doibase
  10.1038/s41534-024-00856-3} {\bibfield  {journal} {\bibinfo  {journal} {npj
  Quantum Information}\ }\textbf {\bibinfo {volume} {10}},\ \bibinfo {pages}
  {64} (\bibinfo {year} {2024})}\BibitemShut {NoStop}%
\bibitem [{\citenamefont {Li}\ \emph {et~al.}(2023)\citenamefont {Li},
  \citenamefont {Yu}, \citenamefont {Wang}, \citenamefont {Xing}, \citenamefont
  {Kong},\ and\ \citenamefont {Zhou}}]{Li2023}%
  \BibitemOpen
  \bibfield  {author} {\bibinfo {author} {\bibfnamefont {Zhi-Hao}\ \bibnamefont
  {Li}}, \bibinfo {author} {\bibfnamefont {Gui-Fang}\ \bibnamefont {Yu}},
  \bibinfo {author} {\bibfnamefont {Ya-Xin}\ \bibnamefont {Wang}}, \bibinfo
  {author} {\bibfnamefont {Ze-Yu}\ \bibnamefont {Xing}}, \bibinfo {author}
  {\bibfnamefont {Ling-Wen}\ \bibnamefont {Kong}}, \ and\ \bibinfo {author}
  {\bibfnamefont {Xiao-Qi}\ \bibnamefont {Zhou}},\ }\bibfield  {title}
  {\enquote {\bibinfo {title} {Experimental demonstration of deterministic
  quantum search algorithms on a programmable silicon photonic chip},}\ }\href
  {\doibase 10.1007/s11433-023-2130-9} {\bibfield  {journal} {\bibinfo
  {journal} {Science China Physics, Mechanics {\&} Astronomy}\ }\textbf
  {\bibinfo {volume} {66}},\ \bibinfo {pages} {290311} (\bibinfo {year}
  {2023})}\BibitemShut {NoStop}%
\bibitem [{\citenamefont {Li}\ \emph {et~al.}(2024{\natexlab{b}})\citenamefont
  {Li}, \citenamefont {Shen}, \citenamefont {Gao},\ and\ \citenamefont
  {Li}}]{Li2024resourceefficient}%
  \BibitemOpen
  \bibfield  {author} {\bibinfo {author} {\bibfnamefont {Xiang}\ \bibnamefont
  {Li}}, \bibinfo {author} {\bibfnamefont {Hanxiang}\ \bibnamefont {Shen}},
  \bibinfo {author} {\bibfnamefont {Weiguo}\ \bibnamefont {Gao}}, \ and\
  \bibinfo {author} {\bibfnamefont {Yingzhou}\ \bibnamefont {Li}},\ }\bibfield
  {title} {\enquote {\bibinfo {title} {Resource {E}fficient {B}oolean
  {F}unction {S}olver on {Q}uantum {C}omputer},}\ }\href {\doibase
  10.22331/q-2024-10-10-1500} {\bibfield  {journal} {\bibinfo  {journal}
  {{Quantum}}\ }\textbf {\bibinfo {volume} {8}},\ \bibinfo {pages} {1500}
  (\bibinfo {year} {2024}{\natexlab{b}})}\BibitemShut {NoStop}%
\bibitem [{\citenamefont {Pokharel}\ and\ \citenamefont
  {Lidar}(2024)}]{Pokharel2024}%
  \BibitemOpen
  \bibfield  {author} {\bibinfo {author} {\bibfnamefont {Bibek}\ \bibnamefont
  {Pokharel}}\ and\ \bibinfo {author} {\bibfnamefont {Daniel~A.}\ \bibnamefont
  {Lidar}},\ }\bibfield  {title} {\enquote {\bibinfo {title}
  {Better-than-classical grover search via quantum error detection and
  suppression},}\ }\href {\doibase 10.1038/s41534-023-00794-6} {\bibfield
  {journal} {\bibinfo  {journal} {npj Quantum Information}\ }\textbf {\bibinfo
  {volume} {10}},\ \bibinfo {pages} {23} (\bibinfo {year} {2024})}\BibitemShut
  {NoStop}%
\bibitem [{\citenamefont {Borish}\ \emph {et~al.}(2020)\citenamefont {Borish},
  \citenamefont {Markovi\ifmmode~\acute{c}\else \'{c}\fi{}}, \citenamefont
  {Hines}, \citenamefont {Rajagopal},\ and\ \citenamefont
  {Schleier-Smith}}]{PhysRevLett.124.063601}%
  \BibitemOpen
  \bibfield  {author} {\bibinfo {author} {\bibfnamefont {V.}~\bibnamefont
  {Borish}}, \bibinfo {author} {\bibfnamefont {O.}~\bibnamefont
  {Markovi\ifmmode~\acute{c}\else \'{c}\fi{}}}, \bibinfo {author}
  {\bibfnamefont {J.~A.}\ \bibnamefont {Hines}}, \bibinfo {author}
  {\bibfnamefont {S.~V.}\ \bibnamefont {Rajagopal}}, \ and\ \bibinfo {author}
  {\bibfnamefont {M.}~\bibnamefont {Schleier-Smith}},\ }\bibfield  {title}
  {\enquote {\bibinfo {title} {Transverse-field ising dynamics in a
  rydberg-dressed atomic gas},}\ }\href {\doibase
  10.1103/PhysRevLett.124.063601} {\bibfield  {journal} {\bibinfo  {journal}
  {Phys. Rev. Lett.}\ }\textbf {\bibinfo {volume} {124}},\ \bibinfo {pages}
  {063601} (\bibinfo {year} {2020})}\BibitemShut {NoStop}%
\bibitem [{\citenamefont {Geier}\ \emph {et~al.}(2021)\citenamefont {Geier},
  \citenamefont {Thaicharoen}, \citenamefont {Hainaut}, \citenamefont {Franz},
  \citenamefont {Salzinger}, \citenamefont {Tebben}, \citenamefont
  {Grimshandl}, \citenamefont {Zürn},\ and\ \citenamefont
  {Weidemüller}}]{doi:10.1126/science.abd9547}%
  \BibitemOpen
  \bibfield  {author} {\bibinfo {author} {\bibfnamefont {Sebastian}\
  \bibnamefont {Geier}}, \bibinfo {author} {\bibfnamefont {Nithiwadee}\
  \bibnamefont {Thaicharoen}}, \bibinfo {author} {\bibfnamefont {Clément}\
  \bibnamefont {Hainaut}}, \bibinfo {author} {\bibfnamefont {Titus}\
  \bibnamefont {Franz}}, \bibinfo {author} {\bibfnamefont {Andre}\ \bibnamefont
  {Salzinger}}, \bibinfo {author} {\bibfnamefont {Annika}\ \bibnamefont
  {Tebben}}, \bibinfo {author} {\bibfnamefont {David}\ \bibnamefont
  {Grimshandl}}, \bibinfo {author} {\bibfnamefont {Gerhard}\ \bibnamefont
  {Zürn}}, \ and\ \bibinfo {author} {\bibfnamefont {Matthias}\ \bibnamefont
  {Weidemüller}},\ }\bibfield  {title} {\enquote {\bibinfo {title} {Floquet
  hamiltonian engineering of an isolated many-body spin system},}\ }\href
  {\doibase 10.1126/science.abd9547} {\bibfield  {journal} {\bibinfo  {journal}
  {Science}\ }\textbf {\bibinfo {volume} {374}},\ \bibinfo {pages} {1149--1152}
  (\bibinfo {year} {2021})}\BibitemShut {NoStop}%
\bibitem [{\citenamefont {Li}\ and\ \citenamefont
  {Shao}(2018)}]{Li2018Unconventional}%
  \BibitemOpen
  \bibfield  {author} {\bibinfo {author} {\bibfnamefont {D.~X.}\ \bibnamefont
  {Li}}\ and\ \bibinfo {author} {\bibfnamefont {X.~Q.}\ \bibnamefont {Shao}},\
  }\bibfield  {title} {\enquote {\bibinfo {title} {Unconventional rydberg
  pumping and applications in quantum information processing},}\ }\href
  {\doibase 10.1103/PhysRevA.98.062338} {\bibfield  {journal} {\bibinfo
  {journal} {Phys. Rev. A}\ }\textbf {\bibinfo {volume} {98}},\ \bibinfo
  {pages} {062338} (\bibinfo {year} {2018})}\BibitemShut {NoStop}%
\bibitem [{\citenamefont {Levine}\ \emph {et~al.}(2018)\citenamefont {Levine},
  \citenamefont {Keesling}, \citenamefont {Omran}, \citenamefont {Bernien},
  \citenamefont {Schwartz}, \citenamefont {Zibrov}, \citenamefont {Endres},
  \citenamefont {Greiner}, \citenamefont {Vuleti\ifmmode~\acute{c}\else
  \'{c}\fi{}},\ and\ \citenamefont {Lukin}}]{PhysRevLett.121.123603}%
  \BibitemOpen
  \bibfield  {author} {\bibinfo {author} {\bibfnamefont {Harry}\ \bibnamefont
  {Levine}}, \bibinfo {author} {\bibfnamefont {Alexander}\ \bibnamefont
  {Keesling}}, \bibinfo {author} {\bibfnamefont {Ahmed}\ \bibnamefont {Omran}},
  \bibinfo {author} {\bibfnamefont {Hannes}\ \bibnamefont {Bernien}}, \bibinfo
  {author} {\bibfnamefont {Sylvain}\ \bibnamefont {Schwartz}}, \bibinfo
  {author} {\bibfnamefont {Alexander~S.}\ \bibnamefont {Zibrov}}, \bibinfo
  {author} {\bibfnamefont {Manuel}\ \bibnamefont {Endres}}, \bibinfo {author}
  {\bibfnamefont {Markus}\ \bibnamefont {Greiner}}, \bibinfo {author}
  {\bibfnamefont {Vladan}\ \bibnamefont {Vuleti\ifmmode~\acute{c}\else
  \'{c}\fi{}}}, \ and\ \bibinfo {author} {\bibfnamefont {Mikhail~D.}\
  \bibnamefont {Lukin}},\ }\bibfield  {title} {\enquote {\bibinfo {title}
  {High-fidelity control and entanglement of rydberg-atom qubits},}\ }\href
  {\doibase 10.1103/PhysRevLett.121.123603} {\bibfield  {journal} {\bibinfo
  {journal} {Phys. Rev. Lett.}\ }\textbf {\bibinfo {volume} {121}},\ \bibinfo
  {pages} {123603} (\bibinfo {year} {2018})}\BibitemShut {NoStop}%
\bibitem [{\citenamefont {Wilk}\ \emph {et~al.}(2010)\citenamefont {Wilk},
  \citenamefont {Ga\"etan}, \citenamefont {Evellin}, \citenamefont {Wolters},
  \citenamefont {Miroshnychenko}, \citenamefont {Grangier},\ and\ \citenamefont
  {Browaeys}}]{PhysRevLett.104.010502}%
  \BibitemOpen
  \bibfield  {author} {\bibinfo {author} {\bibfnamefont {T.}~\bibnamefont
  {Wilk}}, \bibinfo {author} {\bibfnamefont {A.}~\bibnamefont {Ga\"etan}},
  \bibinfo {author} {\bibfnamefont {C.}~\bibnamefont {Evellin}}, \bibinfo
  {author} {\bibfnamefont {J.}~\bibnamefont {Wolters}}, \bibinfo {author}
  {\bibfnamefont {Y.}~\bibnamefont {Miroshnychenko}}, \bibinfo {author}
  {\bibfnamefont {P.}~\bibnamefont {Grangier}}, \ and\ \bibinfo {author}
  {\bibfnamefont {A.}~\bibnamefont {Browaeys}},\ }\bibfield  {title} {\enquote
  {\bibinfo {title} {Entanglement of two individual neutral atoms using rydberg
  blockade},}\ }\href {\doibase 10.1103/PhysRevLett.104.010502} {\bibfield
  {journal} {\bibinfo  {journal} {Phys. Rev. Lett.}\ }\textbf {\bibinfo
  {volume} {104}},\ \bibinfo {pages} {010502} (\bibinfo {year}
  {2010})}\BibitemShut {NoStop}%
\bibitem [{\citenamefont {Bernien}\ \emph {et~al.}(2017)\citenamefont
  {Bernien}, \citenamefont {Schwartz}, \citenamefont {Keesling}, \citenamefont
  {Levine}, \citenamefont {Omran}, \citenamefont {Pichler}, \citenamefont
  {Choi}, \citenamefont {Zibrov}, \citenamefont {Endres}, \citenamefont
  {Greiner}, \citenamefont {Vuleti{\'{c}}},\ and\ \citenamefont
  {Lukin}}]{Bernien2017}%
  \BibitemOpen
  \bibfield  {author} {\bibinfo {author} {\bibfnamefont {Hannes}\ \bibnamefont
  {Bernien}}, \bibinfo {author} {\bibfnamefont {Sylvain}\ \bibnamefont
  {Schwartz}}, \bibinfo {author} {\bibfnamefont {Alexander}\ \bibnamefont
  {Keesling}}, \bibinfo {author} {\bibfnamefont {Harry}\ \bibnamefont
  {Levine}}, \bibinfo {author} {\bibfnamefont {Ahmed}\ \bibnamefont {Omran}},
  \bibinfo {author} {\bibfnamefont {Hannes}\ \bibnamefont {Pichler}}, \bibinfo
  {author} {\bibfnamefont {Soonwon}\ \bibnamefont {Choi}}, \bibinfo {author}
  {\bibfnamefont {Alexander~S.}\ \bibnamefont {Zibrov}}, \bibinfo {author}
  {\bibfnamefont {Manuel}\ \bibnamefont {Endres}}, \bibinfo {author}
  {\bibfnamefont {Markus}\ \bibnamefont {Greiner}}, \bibinfo {author}
  {\bibfnamefont {Vladan}\ \bibnamefont {Vuleti{\'{c}}}}, \ and\ \bibinfo
  {author} {\bibfnamefont {Mikhail~D.}\ \bibnamefont {Lukin}},\ }\bibfield
  {title} {\enquote {\bibinfo {title} {Probing many-body dynamics on a 51-atom
  quantum simulator},}\ }\href {\doibase 10.1038/nature24622} {\bibfield
  {journal} {\bibinfo  {journal} {Nature}\ }\textbf {\bibinfo {volume} {551}},\
  \bibinfo {pages} {579--584} (\bibinfo {year} {2017})}\BibitemShut {NoStop}%
\bibitem [{\citenamefont {Nielsen}(2002)}]{NIELSEN2002249}%
  \BibitemOpen
  \bibfield  {author} {\bibinfo {author} {\bibfnamefont {Michael~A}\
  \bibnamefont {Nielsen}},\ }\bibfield  {title} {\enquote {\bibinfo {title} {A
  simple formula for the average gate fidelity of a quantum dynamical
  operation},}\ }\href {\doibase https://doi.org/10.1016/S0375-9601(02)01272-0}
  {\bibfield  {journal} {\bibinfo  {journal} {Physics Letters A}\ }\textbf
  {\bibinfo {volume} {303}},\ \bibinfo {pages} {249--252} (\bibinfo {year}
  {2002})}\BibitemShut {NoStop}%
\bibitem [{\citenamefont {Yun}\ \emph {et~al.}(2024)\citenamefont {Yun},
  \citenamefont {Wu}, \citenamefont {Yan}, \citenamefont {Jia}, \citenamefont
  {Su},\ and\ \citenamefont {Shan}}]{PhysRevApplied.21.064053}%
  \BibitemOpen
  \bibfield  {author} {\bibinfo {author} {\bibfnamefont {M.-R.}\ \bibnamefont
  {Yun}}, \bibinfo {author} {\bibfnamefont {Jin-Lei}\ \bibnamefont {Wu}},
  \bibinfo {author} {\bibfnamefont {L.-L.}\ \bibnamefont {Yan}}, \bibinfo
  {author} {\bibfnamefont {Yu}~\bibnamefont {Jia}}, \bibinfo {author}
  {\bibfnamefont {Shi-Lei}\ \bibnamefont {Su}}, \ and\ \bibinfo {author}
  {\bibfnamefont {C.-X.}\ \bibnamefont {Shan}},\ }\bibfield  {title} {\enquote
  {\bibinfo {title} {Quantum computation in silicon-vacancy centers based on
  nonadiabatic geometric gates protected by dynamical decoupling},}\ }\href
  {\doibase 10.1103/PhysRevApplied.21.064053} {\bibfield  {journal} {\bibinfo
  {journal} {Phys. Rev. Appl.}\ }\textbf {\bibinfo {volume} {21}},\ \bibinfo
  {pages} {064053} (\bibinfo {year} {2024})}\BibitemShut {NoStop}%
\bibitem [{\citenamefont {Li}\ \emph {et~al.}(2021)\citenamefont {Li},
  \citenamefont {Guo}, \citenamefont {Jin}, \citenamefont {Yan}, \citenamefont
  {Liang},\ and\ \citenamefont {Su}}]{PhysRevA.103.062607}%
  \BibitemOpen
  \bibfield  {author} {\bibinfo {author} {\bibfnamefont {Meng}\ \bibnamefont
  {Li}}, \bibinfo {author} {\bibfnamefont {F.-Q.}\ \bibnamefont {Guo}},
  \bibinfo {author} {\bibfnamefont {Z.}~\bibnamefont {Jin}}, \bibinfo {author}
  {\bibfnamefont {L.-L.}\ \bibnamefont {Yan}}, \bibinfo {author} {\bibfnamefont
  {E.-J.}\ \bibnamefont {Liang}}, \ and\ \bibinfo {author} {\bibfnamefont
  {S.-L.}\ \bibnamefont {Su}},\ }\bibfield  {title} {\enquote {\bibinfo {title}
  {Multiple-qubit controlled unitary quantum gate for rydberg atoms using
  shortcut to adiabaticity and optimized geometric quantum operations},}\
  }\href {\doibase 10.1103/PhysRevA.103.062607} {\bibfield  {journal} {\bibinfo
   {journal} {Phys. Rev. A}\ }\textbf {\bibinfo {volume} {103}},\ \bibinfo
  {pages} {062607} (\bibinfo {year} {2021})}\BibitemShut {NoStop}%
\bibitem [{\citenamefont {Yun}\ \emph {et~al.}(2022)\citenamefont {Yun},
  \citenamefont {Guo}, \citenamefont {Yan}, \citenamefont {Liang},
  \citenamefont {Zhang}, \citenamefont {Su}, \citenamefont {Shan},\ and\
  \citenamefont {Jia}}]{PhysRevA.105.012611}%
  \BibitemOpen
  \bibfield  {author} {\bibinfo {author} {\bibfnamefont {Meng-Ru}\ \bibnamefont
  {Yun}}, \bibinfo {author} {\bibfnamefont {Fu-Qiang}\ \bibnamefont {Guo}},
  \bibinfo {author} {\bibfnamefont {L.-L.}\ \bibnamefont {Yan}}, \bibinfo
  {author} {\bibfnamefont {Erjun}\ \bibnamefont {Liang}}, \bibinfo {author}
  {\bibfnamefont {Y.}~\bibnamefont {Zhang}}, \bibinfo {author} {\bibfnamefont
  {S.-L.}\ \bibnamefont {Su}}, \bibinfo {author} {\bibfnamefont {C.~X.}\
  \bibnamefont {Shan}}, \ and\ \bibinfo {author} {\bibfnamefont
  {Yu}~\bibnamefont {Jia}},\ }\bibfield  {title} {\enquote {\bibinfo {title}
  {Parallel-path implementation of nonadiabatic geometric quantum gates in a
  decoherence-free subspace with nitrogen-vacancy centers},}\ }\href {\doibase
  10.1103/PhysRevA.105.012611} {\bibfield  {journal} {\bibinfo  {journal}
  {Phys. Rev. A}\ }\textbf {\bibinfo {volume} {105}},\ \bibinfo {pages}
  {012611} (\bibinfo {year} {2022})}\BibitemShut {NoStop}%
\bibitem [{\citenamefont {Šibalić}\ \emph {et~al.}(2017)\citenamefont
  {Šibalić}, \citenamefont {Pritchard}, \citenamefont {Adams},\ and\
  \citenamefont {Weatherill}}]{SIBALIC2017319}%
  \BibitemOpen
  \bibfield  {author} {\bibinfo {author} {\bibfnamefont {N.}~\bibnamefont
  {Šibalić}}, \bibinfo {author} {\bibfnamefont {J.D.}\ \bibnamefont
  {Pritchard}}, \bibinfo {author} {\bibfnamefont {C.S.}\ \bibnamefont {Adams}},
  \ and\ \bibinfo {author} {\bibfnamefont {K.J.}\ \bibnamefont {Weatherill}},\
  }\bibfield  {title} {\enquote {\bibinfo {title} {Arc: An open-source library
  for calculating properties of alkali rydberg atoms},}\ }\href {\doibase
  https://doi.org/10.1016/j.cpc.2017.06.015} {\bibfield  {journal} {\bibinfo
  {journal} {Computer Physics Communications}\ }\textbf {\bibinfo {volume}
  {220}},\ \bibinfo {pages} {319--331} (\bibinfo {year} {2017})}\BibitemShut
  {NoStop}%
\bibitem [{\citenamefont {Shi}(2020)}]{PhysRevApplied.13.024008}%
  \BibitemOpen
  \bibfield  {author} {\bibinfo {author} {\bibfnamefont {Xiao-Feng}\
  \bibnamefont {Shi}},\ }\bibfield  {title} {\enquote {\bibinfo {title}
  {Suppressing motional dephasing of ground-rydberg transition for
  high-fidelity quantum control with neutral atoms},}\ }\href {\doibase
  10.1103/PhysRevApplied.13.024008} {\bibfield  {journal} {\bibinfo  {journal}
  {Phys. Rev. Appl.}\ }\textbf {\bibinfo {volume} {13}},\ \bibinfo {pages}
  {024008} (\bibinfo {year} {2020})}\BibitemShut {NoStop}%
\bibitem [{\citenamefont {Lee}\ \emph {et~al.}(2019)\citenamefont {Lee},
  \citenamefont {Kim}, \citenamefont {Jo}, \citenamefont {Song},\ and\
  \citenamefont {Ahn}}]{PhysRevA.99.043404}%
  \BibitemOpen
  \bibfield  {author} {\bibinfo {author} {\bibfnamefont {Woojun}\ \bibnamefont
  {Lee}}, \bibinfo {author} {\bibfnamefont {Minhyuk}\ \bibnamefont {Kim}},
  \bibinfo {author} {\bibfnamefont {Hanlae}\ \bibnamefont {Jo}}, \bibinfo
  {author} {\bibfnamefont {Yunheung}\ \bibnamefont {Song}}, \ and\ \bibinfo
  {author} {\bibfnamefont {Jaewook}\ \bibnamefont {Ahn}},\ }\bibfield  {title}
  {\enquote {\bibinfo {title} {Coherent and dissipative dynamics of entangled
  few-body systems of rydberg atoms},}\ }\href {\doibase
  10.1103/PhysRevA.99.043404} {\bibfield  {journal} {\bibinfo  {journal} {Phys.
  Rev. A}\ }\textbf {\bibinfo {volume} {99}},\ \bibinfo {pages} {043404}
  (\bibinfo {year} {2019})}\BibitemShut {NoStop}%
\bibitem [{\citenamefont {Liu}\ \emph {et~al.}(2022{\natexlab{b}})\citenamefont
  {Liu}, \citenamefont {Shan}, \citenamefont {Yun}, \citenamefont {Wang},
  \citenamefont {Liu}, \citenamefont {Yan}, \citenamefont {Feng},\ and\
  \citenamefont {Su}}]{PhysRevA.106.052610}%
  \BibitemOpen
  \bibfield  {author} {\bibinfo {author} {\bibfnamefont {Bing-Bing}\
  \bibnamefont {Liu}}, \bibinfo {author} {\bibfnamefont {Zheng}\ \bibnamefont
  {Shan}}, \bibinfo {author} {\bibfnamefont {M.-R.}\ \bibnamefont {Yun}},
  \bibinfo {author} {\bibfnamefont {D.-Y.}\ \bibnamefont {Wang}}, \bibinfo
  {author} {\bibfnamefont {B.-J.}\ \bibnamefont {Liu}}, \bibinfo {author}
  {\bibfnamefont {L.-L.}\ \bibnamefont {Yan}}, \bibinfo {author} {\bibfnamefont
  {M.}~\bibnamefont {Feng}}, \ and\ \bibinfo {author} {\bibfnamefont {S.-L.}\
  \bibnamefont {Su}},\ }\bibfield  {title} {\enquote {\bibinfo {title} {Robust
  three-qubit search algorithm in rydberg atoms via geometric control},}\
  }\href {\doibase 10.1103/PhysRevA.106.052610} {\bibfield  {journal} {\bibinfo
   {journal} {Phys. Rev. A}\ }\textbf {\bibinfo {volume} {106}},\ \bibinfo
  {pages} {052610} (\bibinfo {year} {2022}{\natexlab{b}})}\BibitemShut
  {NoStop}%
\end{thebibliography}%
\end{document}